\documentclass{aastex}          %% The manuscript based on AASTeX v5.x
\usepackage{spr-astr-addons}    %% mimicing ASTR journal style
\def\msun{{\rm ~M}_{\odot}}
\def\rsun{{\rm ~R}_{\odot}}

\RequirePackage{color}

\begin{document}

\title{Accretion, Ablation and Propeller Evolution in Close Millisecond Pulsar Binary Systems}

%% Running heads
\shorttitle{Formation and Evolution of Black Widow Pulsars}
\shortauthors{Kiel P.D. \& Taam R.E.}

\author{Paul D. Kiel\altaffilmark{1, 2}}
\email{pdkiel@gmail.com} 
\and
\author{Ronald E. Taam\altaffilmark{1, 3}}
\email{r-taam@northwestern.edu}

\altaffiltext{1}{Center for Interdisciplinary Exploration and Research in Astrophysics (CIERA)
and Dept. of Physics and Astronomy, Northwestern University, 2145 Sheridan Rd, Evanston, IL 60208}
\altaffiltext{2}{Monash affiliate, Monash Centre for Astrophysics, Monash University, Victoria 3800, Australia}
\altaffiltext{3}{Academia Sinica Institute of Astronomy and Astrophysics - TIARA,
P.O. Box 23-141, Taipei, 10617 Taiwan}

\begin{abstract}
A model for the formation and evolution of binary millisecond radio pulsars in systems with 
low mass companions ($< 0.1 \msun$) is investigated using a binary population synthesis technique. 
Taking into account the non conservative evolution of the system due to mass loss from 
an accretion disk as a result of propeller action and from the companion via ablation by the pulsar, 
the transition from the accretion powered to rotation powered phase is investigated.  It is shown 
that the operation of the propeller and ablation mechanisms can be responsible for the 
formation and evolution of black widow millisecond pulsar systems from the low mass X-ray binary 
phase at an orbital period of $\sim 0.1~$day.  For a range of population synthesis input 
parameters, the results reveal that a population of black widow millisecond pulsars characterized 
by orbital periods as long as $\sim 0.4$ days and companion masses as low as $\sim 0.005 \msun$ 
can be produced.  The orbital periods and minimum companion mass of this radio millisecond pulsar 
population critically depend on the thermal bloating of the semi-degenerate hydrogen mass losing 
component, with longer orbital periods for a greater degree of bloating.  Provided that 
the radius of the companion is increased by about a factor of $2$ relative to a fully 
degenerate, zero temperature configuration, an approximate agreement between observed long orbital 
periods and theoretical modeling of hydrogen rich donors can be achieved.  We find no discrepancy 
between the estimated birth rates for LMXBs and black widow systems, which on average are 
$\sim 1.3\times10^{-5}~{\rm yr}^{-1}$ and $1.3\times10^{-7}~{\rm yr}^{-1}$ respectively.
\end{abstract}

\keywords{binaries: close --- magnetic fields --- pulsars: general --- pulsars: individual (PSR 
J1023+0038)  --- stars: neutron}

\section{Introduction}

The discovery of a radio pulsar with a $1.69~{\rm ms}$ spin period in J102347.67+003841.2 (Archibald 
et al. 2009), a system characterized by a 4.75 hr binary orbital period (Woudt et al. 2004), has 
provided observational confirmation of the link between rotation powered radio millisecond pulsars 
(MSPs) and the low mass X-ray binary (LMXB) phase.  This source is of special interest since it 
appears to have undergone a transformation from a LMXB to a recycled MSP (see Bond et al. 2002; 
Thorstensen \& Armstrong 2005; Wang et al. 2009). Takata, Cheng, \& Taam (2010) suggested that 
the emission of $\gamma$-rays from the pulsar magnetosphere was important in facilitating the 
transformation of a MSP from the accretion powered to the rotation powered phase.

Among the binary MSPs, there is a class of systems characterized by short orbital periods ($< 1$ day),
of which J102347.67+003841.2 is a member, from which we define two distinct populations. In one 
population, known as ultra compact binary X-ray MSPs (UCXBs), systems consist of MSP-white 
dwarf (WD) components with orbital periods $P_{\rm orb} <0.07~{\rm days}$ and companion masses, 
$M_{\rm c}<0.02 \msun$. The UCXB progenitors are thought to have experienced two common envelope (CE) 
phases, one initiated by the neutron star (NS) progenitor -- while the companion was still on the 
main sequence (MS) -- and the other initiated by a helium or carbon oxygen WD progenitor. The other 
population, known as black widow binaries, is composed of a radio MSP and brown dwarf (partially 
degenerate MS star) with $0.1<P_{\rm orb} <1.0~{\rm days}$ and $M_{\rm c} < 0.07 \msun$.  In contrast 
to the former population, these MSP binaries display regular eclipses of the pulsed signal over a 
portion of their orbit, providing direct evidence that the companion is undergoing ablation by the 
MSP.  One of the primary differences between the black widow systems and the UCXB is related 
to the size of the companion star.  As inferred from observations (King et al. 2005), the stellar 
radii of the companions are $\sim 5$ times smaller in UCXB systems, indicating that thermal 
bloating of the donor is important (Nelson \& Rappaport 2003; King et al. 2005). Here, the 
donor refers to the star that is overflowing its Roche lobe.  This difference in radii results in 
contrasting orbital periods, for the same companion mass, between these two populations.

In this paper we explore the evolution of a system from the LMXB phase to the binary MSP phase via 
a binary population synthesis technique taking into account the influence of propeller action in 
ejecting mass from an accretion disk and via ablation from the companion by a pulsar.  Attention 
is focused on examining the long term orbital evolution of compact binary MSPs and black widow 
pulsars when including thermal bloating, ablation and propeller evolution into our models. In 
particular, we show that the transition from the LMXBs to black widow radio pulsars takes place when 
the models account for the propeller mechanism and ablation of the NS companion.

The population synthesis technique allows us to identify important and possibly interesting 
evolutionary scenarios and pathways in a simple manner -- this is one of the goals of the work 
presented here.  Hence, the population synthesis method is preferred over the use of a detailed 
code because it facilitates the modelling of many millions of different systems in exploring 
the influence of numerous assumptions and parameter values.  Detailed codes, although more 
accurate in their physical output, require significantly greater computational expense.

In the next section, we outline the main assumptions for modeling accretion, ablation and 
propeller action in our population synthesis study and describe our method. 
The numerical results 
are presented and compared to detailed simulations in \S3.  Example outcomes for a range of 
systems are described in \S4 with specific application to the black widow pulsar formation rates in 
\S5.  A discussion of our results is presented in \S6, and we conclude in \S7.
 
\section{Formulation and Methodology}

As mass loss from the system due to propeller action and/or from ablation of the companion plays an 
important role in the evolution considered here, the modeling of these effects is described below. 
 
\subsection{Modeling accretion, ablation and propeller physics}
\label{s:modelling}

Mass transfer from one star to another greatly affects both the stellar components and the 
orbital evolution of the binary system. Here, we outline the main influence on the stellar 
components and orbit, focusing on the evolution of the stellar spins and the orbital angular 
momentum.  Since the evolution during mass transfer depends upon the evolutionary phase of the 
system, we distinguish three regimes corresponding to the standard mass transfer, ablation, and 
propeller phases.

In the semi-detached phase, the amount of mass transferred, $\Delta M$, mass accreted, $\Delta 
M_{\rm a}$, and mass lost from the system, $\Delta M_{\rm lost}$, is calculated according to 
Hurley, Tout \& Pols (2002), where $\Delta M = \Delta M_{\rm lost} + \Delta M_{\rm a}$. If the 
mass transfer rate is greater than the Eddington rate, then $\Delta M_{\rm lost} > 0$.  The Eddington 
mass transfer rate, above which some mass is assumed lost from the system, is $\dot{M}_{\rm Edd} 
= 2.08\times 10^{-3} \left( 1+X \right) R_{\rm a}\msun~{\rm yr}^{-1}$, where $X = 0.76 - 3.0Z$ is 
the hydrogen abundance, $Z$ is the metallicity and $R_{\rm a}$ is the neutron star radius in solar 
units.  Note, subscript `a' denotes accretor parameters (where the accretor is the NS), while 
subscript `d' denotes donor star parameters (see below).
  
\subsubsection{Standard RLOF evolution}
\label{s:rlof}
We describe standard Roche lobe overflow (RLOF) evolution when the operation of the propeller or 
ablation mechanism (see below) is unimportant.  The removal of $\Delta M$ from the donor modifies 
the donors spin which, in turn, is coupled to the orbit. Assuming that the system is in a steady 
state, the change in spin angular momentum is given by 
\begin{equation}
\left(\Delta J_{\rm d~spin}\right)_{\rm new} = \Delta J_{\rm d~spin} - \Delta M R_{\rm d}^2 
\Omega_{\rm d},
\end{equation}
and the change in orbital angular momentum by 
\begin{equation}
\left(\Delta J_{\rm orb}\right)_{\rm new} = \Delta J_{\rm orb} + f_{\rm c}\Delta M R_{\rm d}^2 
\Omega_{\rm d}.
\end{equation}
Here, $R_{\rm d}$ is the radius of the donor, $\Omega_{\rm d}$ is its angular velocity and 
$f_{\rm c}$ is a coupling factor. 
$\Delta J_{\rm d~spin}$ and $\Delta J_{\rm orb}$ contain 
other evolutionary modifications to the donor spin and orbital angular momentum in the same 
time step as our modifications owing to mass transfer.  These additional adjustments are associated
with tidal torques, magnetic braking (see Eqn.~\ref{e:mb}) and gravitational radiation (see 
Eqns.~\ref{e:grJ} and \ref{e:gre}).
The coupling factor indicates the amount of angular momentum 
that is introduced to the orbit.  For simplicity we assume $f_{\rm c} = 1$ unless otherwise stated.  
If the accretion rate onto the NS is super-Eddington, then 
\begin{equation}
\left(\Delta J_{\rm orb}\right)_{\rm new} = \Delta J_{\rm orb} - \frac{\Delta M_{\rm lost} M_{\rm d}^2}
{M_{\rm tot}^2} a^2\Omega_{\rm orb}\sqrt{1-e^2} ,
\end{equation}
where $\Omega_{\rm orb}$ is the orbital angular velocity, $e$ is the orbit eccentricity, $a$ is the 
orbital separation, and $M_{\rm tot} = M_{\rm a} + M_{\rm d}$ is the total mass of the system.  
The spin of the accretor is updated from 
\begin{equation}
\left(\Delta J_{\rm a~spin}\right)_{\rm new} = \Delta J_{\rm a~spin} + \Delta M_{\rm a} \sqrt{GM_{\rm 
a} R_{\rm in}},
\end{equation}
while the change in orbital angular momentum is updated by 
\begin{equation}
\left(\Delta J_{\rm orb}\right)_{\rm new} = \Delta J_{\rm orb} - \Delta M_{\rm a} \sqrt{M_{\rm a} 
G R_{\rm in}}.
\end{equation}
The lever arm $R_{\rm in}$ is set equal to the larger of the neutron star radius or the 
magnetospheric radius, 
i.e., $R_{\rm in} = MAX(R_{\rm a},R_{\rm m})$, which is given as $R_{\rm m} = 1.7\times10^{-4} 
R_\odot ~B_{\rm s}^{4/7}R_{\rm NS}^{12/7}M_{\rm NS}^{-1/7} \dot{M}^{-2/7}$.  Here, $R_{\rm m}$ 
is taken to be half of the Alfven radius, which corresponds to the location where the magnetic 
field pressure equals the ram pressure of in falling material (Frank, King \& Raine 2002).  The 
pulsar magnetic field strength, $B_{\rm s}$, is in units of Gauss, $R_{\rm NS}$ and $M_{\rm NS}$ 
are the pulsar radius and mass respectively in solar units, and $\dot{M}$ is the mass transfer rate 
in $M_\odot {\rm yr}^{-1}$.

\subsubsection{Propeller, spin down and equilibrium spin down modes}
The concept of propeller action operating in interacting binary systems was first suggested as 
a mechanism to facilitate mass loss in a pioneering study by Illarionov \& Sunyaev (1975).  Such an
evolution requires a NS with a sufficiently strong magnetic field and sufficient rotation, at the 
magnetospheric boundary, to halt accretion and eject material from the inner edge of the accretion 
disk.  In this process, the propeller mechanism expels incoming material at the expense of the NS's 
angular momentum.  During this phase the NS spin angular momentum and orbital angular momentum are 
updated as described by Kiel et al. (2008) and outlined below.
 
In this case, the donor spins down as $\Delta M$ is removed, coupling it to the orbit as described 
above.  This $\Delta M$, however, is now removed from the system so that the change in orbital 
angular momentum is, 
\begin{equation}
\label{e:proporb}
\left(\Delta J_{\rm orb}\right)_{\rm new} = \Delta J_{\rm orb} - \Omega_{\rm K}(R_{\rm m}) 
R_{\rm m}^2 \Delta M,
\end{equation}
where $\Omega_{\rm K}(R_{\rm m})$ is the Keplerian angular velocity at the magnetospheric radius.  
The accretor spins down according to 
\begin{equation}
\left(\Delta J_{\rm a~spin}\right)_{\rm new} = \Delta J_{\rm a~spin} - 
\left( \Omega_{\rm K}(R_{\rm m}) - \Omega_{\rm a} \right) R_{\rm m}^2 \Delta M.  
\label{e:propspin}
\end{equation}
We limit the time step so that the second term on the right side of Eqn.~\ref{e:propspin} does not
introduce numerical instabilities.

Through accretion and propeller evolution an equilibrium spin period can be achieved, which produces 
a `spin up line' in the pulsar spin period-magnetic field distribution.  Following the magnetic 
dipole model of Arzoumanian, Cordes, Wasserman (1999, equ. 2) we adopt the spin up line given by 
$B_{\rm eq} = f_{\rm eq}1.025\times 10^{12}P^{7/6}$, where the spin period, $P$, is expressed in 
seconds, the critical equilibrium magnetic field, $B_{\rm eq}$, is in Gauss and $f_{\rm eq}$ is a 
parameter encapsulating the uncertainties in the relation and is of order unity.  In this picture, 
the pulsar spins up when $B<B_{\rm eq}$, spins down when $B>B_{\rm eq}$, and is held at the spin 
equilibrium period when $B \sim B_{\rm eq}$.  To prevent the system from alternating between these 
two states in successive time steps, when $B\sim B_{\rm eq}$, some fraction (typically half) of the 
mass is allowed to accrete during that time step with a concomitant decrease in the field (see 
\S\ref{s:method}).  The pulsar is now forced to rotate at the spin equilibrium of the new pulsar 
magnetic field, and the pulsar evolves along the spin up line to shorter spin periods provided that 
equilibrium can be maintained.  During the equilibrium phase, the rotation of the accretor is coupled 
to the orbit similar to Eqn.~\ref{e:proporb}.  If ablation occurs during the operation of the 
propeller mechanism (either spin down or the equilibrium scenario), we ignore the propeller mechanism 
and instead implement the ablation mechanism as described below (because the matter is ejected before 
arriving at the magnetosphere of the NS).  In addition, the accreting NS is not allowed to spin up 
beyond the spin equilibrium line.

\subsubsection{Ablation evolution}
Ablation of the companion can occur if the rate of energy deposition into its envelope is 
sufficiently high.  This can result from irradiation of the companion by the pulsar as a 
consequence of the pulsar electromagnetic radiation and/or its relativistic wind.  For a donor 
star sufficiently close to its NS companion, matter can be lifted off its surface and lost from 
the system.  During the ablation phase the rate of mass loss from the companion, 
$\dot{M}_{\rm ablated}$, follows from the simple prescription of van den Heuvel \& van Paradijs 
(1988) such that, 
\begin{equation}
\dot{M}_{\rm ablated} =  f_{\rm e}\left(\frac{2}{V_{\rm esc}^2}\right) \left(\frac{R_{\rm d}}{2a}\right)^2 
\frac{2R_{\rm a}^6B_{\rm s}^2}{3c^3} \left(\frac{2\pi}{P}\right)^4
\end{equation}
Here $V_{\rm esc}$ is the escape velocity of the companion and $f_{\rm e}$ is an efficiency factor 
for converting the pulsar luminosity into mass loss.  As this factor is not well determined, it is 
treated as a parameter, which unless otherwise stated is taken to be $0.1\%$.  We note that a 
value of $0.1\%$ is a conservative choice in comparison to estimates of the ablation efficiency 
adopted by van den Heuvel \& van Paradijs (1988), Phinney et al. (1988), and Ruderman, Shaham \& 
Tavani (1989).  When the donor is a giant star we limit the mass ablated in one time step, $\Delta 
t$, to be no more than the mass of the envelope -- in practice this only becomes important when the 
majority of the envelope has already been lost.  For completeness, we also require $\dot{M}_{\rm 
ablated} < M_{\rm d}/\Delta t$.  The donor mass is updated accounting for both the RLOF mass 
loss and the ablation mass loss.  For the latter, the material removed from the orbit is described 
as a wind from the companion.  The change in orbital angular momentum is given by 
\begin{eqnarray}
\label{e:ab_j}
\left(\Delta J_{\rm orb}\right)_{\rm new} = \Delta J_{\rm orb} - \frac{(\Delta M_{\rm ablated})M_{\rm 
a}^2}{M_{\rm tot}^2}a^2\Omega_{\rm orb}\sqrt{1-e^2}.
\end{eqnarray}
The specific angular momentum of the matter lost from the system due to the donor is 
$M_{\rm a}/M_{\rm d}$ times the specific angular momentum of the binary system. This loss 
coupled with the mass loss from the system due to ablation of the companion will tend to lead 
to an increase in orbital period.  The corresponding change in the spin of the star is,
\begin{equation}
\left(\Delta J_{\rm d~spin}\right)_{\rm new} = \Delta J_{\rm d~spin} - \Delta M_{\rm ablated} R_{\rm d}^2 \Omega_{\rm d}.
\end{equation}
The ablation mechanism is allowed to occur even when the companion does not fill its Roche lobe, 
as long as the pulsar is sufficiently powerful. 

\subsection{Distinguishing between accretion, ablation and propeller}
\label{s:distinguish}

The regimes distinguishing the phases of accretion, ablation and propeller can be delineated by 
comparison of the light cylinder radius, magnetospheric radius, and the corotation radius. The light 
cylinder radius, $R_{\rm lc} = cP/2\pi$, defines the radius at which matter coupled to the magnetic 
field rotates at the speed of light, c.  The magnetospheric radius, is defined in \S\ref{s:rlof}.
The co-rotation radius, $R_{\rm co} = \left( GM_{\rm NS}P^2 /4\pi^2 \right)^{1/3}$, represents the 
radius for which the Keplerian angular velocity in the accretion disk is equal to the NS angular 
velocity.  For cases where the magnetosphere breaches the light cylinder ($R_{\rm m} > R_{\rm lc}$), 
the magnetospheric emission in the pulsar may be activated facilitating ablation.  On the other hand, 
propeller action can be enabled if $R_{\rm m} > R_{\rm co}$. 

Analogous to the magnetospheric radius, which is important in determining the evolutionary phases 
of the system, there exists critical mass accretion rates that determine the evolution in our study.  
We adopt the critical accretion rate, $\dot{M}_{\gamma \rm{c}}$, derived by Takata, Cheng \& Taam 
(2010; their equ. 10), 
\begin{equation}
\dot{M}_{\gamma \rm{C}} = 3.3\times 10^{-12} 
P_{-3}^{1/2}B_8^{1/2}s_1^{1/2}R_6^{3/2}E_{0.1}^{-2}M_{1.4}^{-1}~{\msun yr^{-1}},
\label{e:gammamdot}
\end{equation}
where the pulsar may halt accretion.
Here, $P_{-3}$ is the NS rotational period in units of $10^{-3}~$s, $B_8$ is the NS magnetic 
field in units of $10^8$G, $s_1 $ is the ratio of curvature radius of the magnetic field line to 
the light cylinder radius and assumed to be unity, $R_6$ is the NS radius in $10^6$cm, $E_{0.1}$ 
is the X-ray photon energy in units of $100~{\rm eV}$, and $M_{1.4}$ is the NS mass in units of 
$1.4\msun$.

An additional critical rate is associated with the stability of the accretion disk since we assume 
that the pulsar can turn on during the quiescent state of the system when it exhibits X-ray transient 
phenomena (see below).  The latter critical accretion rate depends upon the composition of the disk 
and on the degree of irradiation (Lasota, Dubus \& Kruk 2008).  For a helium (He) star or He WD we 
use the He disk models, whereas H disk models are used for MS or giant star companions.  For 
simplicity we only consider the irradiated disk models (Lasota, Dubus \& Kruk 2008) and assume that 
the outer edge of the disk extends to 0.8 times the NS Roche radius, however, we included the 
non-irradiated case as an option within \textsc{BSE}.  The pure hydrogen equation is (Lasota, 
Dubus \& Kruk 2008; Appendix A),
%\begin{equation}
%\dot{M}_{\rm non-irr} = 2.64\times 10^{15} R_{10}^{2.58}M_1^{-0.85}~{\rm g~s^{-1}}
%\label{e:lasota1}
%\end{equation}
%and
\begin{equation}
\dot{M}_{\rm irr} = 1.5\times 10^{-11} R_{10}^{2.39}M_1^{-0.64}~{\msun~yr^{-1}},
\label{e:lasota2}
\end{equation}
and the helium disk equation is,
\begin{equation}
\dot{M}_{\rm irr} = 3.3\times 10^{-10} R_{10}^{2.51}M_1^{-0.74}~{\msun~yr^{-1}}.
\label{e:lasota3}
\end{equation}
Here $R_{10}$ is the outer disk radius in units of $10^{10}$cm and $M_1$ is the NS 
mass in solar units.
 
Given the radii defined above, there exist four regimes of importance for the evolution of the 
system.  {\bf (1)} In cases for which $R_{\rm m} < R_{\rm co}$ and $R_{\rm m} < R_{\rm 
lc}$ accretion occurs since neither ablation nor propeller evolution develops. {\bf (2)} If 
$R_{\rm m} > R_{\rm co}$ and $R_{\rm m} < R_{\rm lc}$, the pulsar can not ablate the companion, 
but the propeller mechanism can operate.  {\bf (3)} When $R_{\rm m} < R_{\rm co}$ and $R_{\rm m} > 
R_{\rm lc}$ it is possible for the pulsar to ablate its companion without the action of the 
propeller phase taking place. For mass accretion regimes in which the accretion disk is unstable 
($\dot{M} < \dot{M}_{\rm irr}$), we assume the pulsar is activated during the first quiescent 
phase.  If the gamma-ray irradiation mechanism is effective ($\dot{M} < \dot{M}_{\gamma {\rm c}}$) 
accretion is halted and ablation occurs.  {\bf (4)} Finally, both propeller and ablation can occur 
if $R_{\rm m} > R_{\rm co}$ and $R_{\rm m} > R_{\rm lc}$.  To avoid unnecessary complication, we 
assume that during the ablation phase of the MSP companion, material is directly ejected from the system 
and not transferred to the disk.

\subsection{Population synthesis method}
\label{s:method}

The binary population synthesis is performed using \textsc{binpop}, a package developed in Kiel et 
al. (2008).  \textsc{binpop} employs a rapid Binary Stellar Evolution (\textsc{bse}) algorithm as 
described in Hurley, Tout \& Pols (2002)\footnote{Freely accessible at 
http://astronomy.swin.edu.au/$\sim$jhurley/} and Hurley, Pols \& Tout (2000), with updates 
provided in Kiel \& Hurley (2006), Kiel et al. (2008) and Kiel \& Hurley (2009).  
An important addition to \textsc{bse}, for this work, was the inclusion of pulsar physics in both isolation
and binary systems (Kiel at al. 2008), allowing the user to follow pulsar magnetic braking and 
magnetic field decay (Ostriker \& Gunn 1969), accretion induced field decay and spin up, propeller 
evolution, pulsar death lines and electron capture SNe.

Each system, initially comprised of two MS stars, is evolved from a random birth age, 
between $0-T_{\rm max}$, to the assumed age of the Galaxy, $T_{\rm max} = 12~{\rm Gyr}$.  
This leads to an assumed time independent star formation rate, 
the magnitude of which can be scaled when calculating formation rates (see \S~\ref{s:rates}).
The initial binary values and standard evolution assumptions used here closely follow Table~1 of 
Kiel, Hurley, \& Bailes (2010), unless otherwise stated. 
 
For our base model (Model A) accretion induced field decay is chosen to follow an inverse function 
(equ.~9 of Kiel et al. 2008) with a scaling factor of $M_\star = 10^{-4}\msun$ (Shibazaki et 
al. 1989) and a pulsar magnetic field decay timescale of $900~{\rm Myr}$.  This form of decay is 
adopted as the overall shape of the theoretical pulsar distribution better reproduces the observed 
distribution of pulsars in the spin period-spin period derivative diagram than equ.~8 of Kiel et al. 
(2008).  Three supernova Maxwellian kick distributions are adopted which correspond to core 
collapse (with a dispersion of $\sigma_{\rm CC} = 265~{\rm km~s}^{-1}$), accretion induced 
collapse ($\sigma_{\rm AIC} = 90~{\rm km~s}^{-1}$), and electron capture (as defined by Kiel et al. 
2008; $\sigma_{\rm ECS} = 90~{\rm km~s}^{-1}$).

LMXBs have smaller orbital separations than the radii of their compact object progenitor.  A solution 
was suggested in Paczynski (1976), where two compact cores (in this case a compact core can be a 
MS star) spiraled in towards each other within a common envelope.  Friction from the stellar passage 
within the envelope is assumed to drive off the envelope at the expense of orbital energy.  One of 
the compact objects was the core of a giant whose envelope supplied the common envelope now shared 
by the two cores.  The other core can be a highly evolved star (BH, NS or WD) or a MS star.

In LMXB formation the entire envelope needs to be removed and a surplus of orbital angular momentum 
to remain if a close binary is to exist.  This simple scenario leads to an energy balance equation 
of CE evolution, of which the uncertainties are encapsulated into two parameters, the CE efficiency 
parameter, $\alpha_{\rm CE}$, and the stellar structure parameter, $\lambda$, related to the 
binding energy of the CE.  Much uncertainty surrounds the evaluation of $\alpha_{\rm CE}$ (Tutukov 
\& Yungelson 1996; Kiel \& Hurley 2006), however, typically $\alpha_{\rm CE} = 1$ (Willems \& Kolb 
2002; Pfahl et al. 2003; Voss \& Tauris 2003; Kiel \& Hurley 2006), $\alpha_{\rm CE} \lambda = 1$ 
(Belczynski et al. 2002; Nelemans \& Tout 2005; Pfahl, Podsiadlowski \& Rappaport 2005; Belczynski 
et al. 2008) or $\alpha_{\rm CE} = 3$ (Kiel \& Hurley 2006; Kiel et al. 2008; Kiel, Hurley \& Bailes 
2010; Hurley et al. 2010). 
 
The treatment of the CE process itself varies and can result in different values of $\alpha_{\rm CE}$
corresponding to equivalent efficiencies.  For example, in \textsc{bse} the binding energy to be 
driven off is that of the giant star prior to the CE event,
\begin{equation}
E_{\rm bind} = -\frac{G M \left( M-M_{\rm c}\right)}{\lambda R},
\end{equation}
where $G$ is the gravitational constant, $M$ is the donor mass, $M_{\rm c}$ is the donor core mass,
$R$ is the donor radius and $\lambda$ is the donor stellar structure parameter.  This binding energy 
is balanced by the difference between the initial orbital energy of the binary,
\begin{equation}
E_{orb,i} = \frac{-GM_{\rm c}m}{2a_{\rm i}}
\end{equation}
and the final orbital energy,
\begin{equation}
E_{orb,f} = \frac{-GM_{\rm c}m}{2a_{\rm f}}
\end{equation}
such that,
\begin{equation}
\frac{M \left( M-M_{\rm c}\right)}{\lambda R} = \frac{\alpha_{\rm CE}M_{\rm c}m}{2}\left( \frac{1}{a_{\rm f}} - \frac{1}{a_{\rm i}}\right)
\end{equation}
Here $m$ is the compact object, $a_{\rm i}$ is the initial orbital separation at the onset of CE and 
$a_{\rm f}$ is the final orbital separation once the envelope has been entirely removed. An 
alternative formulation of the CE prescription takes the binding energy between the envelope mass 
and the combined mass of the two compact objects (Iben \& Livio 1993; Yungelson et al. 1994),
\begin{equation}
E_{\rm bind} = -\frac{G \left(M+m\right)\left( M-M_{\rm c}\right)}{\lambda R},
\end{equation}
or the entire donor mass is used in the initial orbital equation (Webbink 1984;
de Kool 1990; Podsiadlowski et al. 2003; Belczynski et al. 2008),
\begin{equation}
E_{orb,i} = \frac{-GMm}{2a_{\rm i}}.
\end{equation}
In this latter prescription a value of $\alpha_{\rm CE} = 1$ is equivalent in efficiency to a value 
of $\alpha_{\rm CE}\sim 3$ in the \textsc{bse} method (see Kiel \& Hurley 2006; Hurley et al. 2010).

The stellar structure parameter is another source of uncertainty in the CE prescription.  Although 
the value is known to vary with stellar type and age, many previous models have assumed a constant 
value, typically $\lambda = 0.5$ (Portegies Zwart \& Yungelson 1998; Hurley, Tout \& Pols 2002; 
Belczynski et al. 2002; Podsiadlowski, Rappaport \& Han 2003). The variation with stellar 
age has also been examined in detailed stellar evolution codes (Dewi \& Tauris 2001; Ivanova \& 
Taam 2003; Podsiadlowski et al. 2003), where $\lambda$ is shown to vary between $0.02-0.7$.
Population synthesis calculations that use a non constant value of $\lambda$ are Voss \& Tauris 2003; 
Podsiadlowski et al. 2003; Kiel \& Hurley 2006; Kiel et al. 2008, 2010; Hurley et al. 2010).
Since its inception \textsc{bse} has been updated to include an algorithm that calculates 
a value of $\lambda$ based on comparison to the detailed models of Pols et al. (1998), 
see Kiel \& Hurley (2006) and Hurley et al. (2010).

Detailed treatment of the CE process with hydrodynamic models has been attempted (Bodenheimer 
\& Taam 1984; Taam \& Sandquist 2000; Ricker \& Taam 2008; Podsiadlowski et al. 2010), but has yet 
to produce a cohesive physical understanding of the entire process.  Podiadlowski et al. (2010) 
present an intriguing scenario which shows the simplicity of the $\alpha$-formalism (Nelmans \& Tout 
2005).
  
The evolution of short period systems is determined by the angular momentum losses associated with 
magnetic braking of the companion (Schatzman 1962).  When tidally coupled the spin down of the star 
owing to magnetic braking removes angular momentum from the orbit to maintain the spin-tidal coupling.
We use Eqn.~50 of Hurley, Tout \& Pols (2002), which corresponds closely (within a factor of $3$) to 
the prescription developed by Verbunt \& Zwaan (1981) as 
\begin{equation}
\label{e:mb}
\dot{J}_{\rm mb} = -5.83\times10^{-16}\frac{M_{\rm d,env}}{M_{\rm d}}\left( R_{\rm d} \Omega_{\rm d,spin} \right)^3 \msun \rsun~{\rm yr}^{-2}.
\end{equation}
Here, $M_{\rm d,env}$ is the mass in the envelope of the donor (in this case, for low-mass MS stars 
the envelope mass is the entire stellar mass) and $\Omega_{\rm d,spin}$ is the donor spin rate.  We 
note that eqn.~\ref{e:mb} is just one of several forms that has been proposed over the years as other 
forms are given by Chaboyer, Demarque \& Pinsonneault (1995), Andronov et al. (2003) and Ivanova \& 
Taam (2003).  In addition to angular momentum losses associated with magnetic braking, gravitational 
radiation also removes angular momentum from the orbit.  Here, we make use of equations~48-49 from 
Hurley, Tout \& Pols (2002), which under the assumption of the weak field approximation gives,
\begin{equation}
\label{e:grJ}
\dot{J}_{\rm gr} = K_{\rm gr} J_{\rm orb} \left( 1+7/8e^2 \right)
\end{equation}
and
\begin{equation}
\label{e:gre}
\dot{e} = K_{\rm gr} e \left( 19/6 + 121/96e^2 \right)
\end{equation}
where $K_{\rm gr} = -8.315\times10^{-10}\frac{M_{\rm d}M_{\rm a}M_{\rm tot}}{a^{4}\left(1-e^2 \right)^{5/2}}$.

\subsubsection{The companion}

When a MS star loses sufficient mass, hydrogen burning cannot be maintained, and the stellar 
structure changes from a low mass, fully convective, star to a star supported by electron degeneracy 
pressure. 
The degenerate configuration is approximated by a $R \propto M^{-1/3}$ power law, and for the 
simplest of assumptions (zero temperature) the radius is roughly a factor of $2.5 \sim (1+X)^{5/3}$ 
times larger than a He WD of the same mass.  However, the temporal history of the mass transfer and 
binary evolution up to this transition leads to departures from this simple description.  It should 
be pointed out that within \textsc{bse} the properties of the mass transferring components are 
calculated from fits to detailed models of stars in thermal equilibrium.  In addition, the low mass 
MS like stars in a state of mass transfer originating from stars more massive than $1 \msun$ do not 
account for deviations of the central hydrogen content associated with nuclear evolution which lead 
to shorter orbital periods than obtained by using a mass-radius relation given by Eqn.~\ref{e:m-r}.

For the low mass short orbital period binaries the thermal timescale of the donor is long compared 
to the mass loss timescale.  For example the Kelvin-Helmholtz timescale for a MS star near the 
orbital period minimum, corresponding to a mass $M\sim0.07\msun$, is $\tau_{\rm 
KH}\sim 4\times10^{9}~$yr, which increases with decreasing mass.  On the other hand, the mass 
transfer time scale for systems near the minimum orbital period are $<2\times10^9~$yr.  Therefore, 
these stars will be out of thermal equilibrium and not fully degenerate.  This departure from 
thermal equilibrium results in so called thermal bloating for which the radius of the star is larger 
than in a fully degenerate state.  Although it is desirable to include the affect of thermal 
bloating in detail in our synthesis, we follow the parameterization of the mass radius relation 
governing low mass hydrogen stars adopted in \textsc{bse} (see Eqn. 24 in Hurley, Tout \& Pols 2002), 
so that the partially degenerate MS radius is, 
\begin{equation}
R_{\rm MS} = 0.0128 \left(1+X \right)^{5/3} k \left( \frac{M_{\rm d}}{M_\odot} \right)^{-1/3}\rsun.
\label{e:m-r}
\end{equation}

We note that Hurley, Tout \& Pols (2002), following Tout et al. (1997), implicitly assumed these 
stars to be out of thermal equilibrium and that $k=2$.  The form of this equation follows Tout et al. 
(1997), and is parameterized in a similar manner by Nelson \& Rappaport (2003) and King et al. (2005) 
in their semi analytical approaches.  Here, $k$, is the bloating factor corresponding to the ratio 
of the radius to its fully degenerate zero temperature radius.  Owing to the very simple prescription 
based method of modeling here, we do not capture this effect in a fully consistent way, but expect 
that the trends produced in the evolutions are not severely affected.  The distended behavior of 
the companion star has been noted in previous studies in which similar systems were evolved in 
detail (Nelson \& Rappaport 2003; Rappaport, Joss \& Webbink 1982; Paczynski \& Sienkiewicz 1981). 
In particular, Rappaport, Joss \& Webbink (1982) and Nelson \& Rappaport (2003) show clearly that 
an increase in angular momentum losses (shorter mass loss time scales) results in more distended 
companion stars.
  
In addition to mechanisms related to pulsar properties in driving the evolution of the system, 
dynamical mass transfer leading to a merger of the binary components can also take place.
Following Ruderman \& Shaham (1985) we assume the mass transfer from He, carbon-oxygen (CO) or 
oxygen-neon (ONe) WDs becomes dynamically unstable for some critical mass ratio, assumed to be 
$\sim0.006$ (which for $M_{\rm NS}\sim1.35 \msun$ gives $M_{\rm c}\sim0.008 \msun$). 
This critical ratio is lower than the value assumed by Ruderman \& Shaham (1985), but 
was chosen to match well with the observed minimum mass of ultra-compact MSP binaries 
(see \S~\ref{s:evex}).  This evolutionary phase has not been established in the general context, 
and we assume hydrogen-rich semi-degenerate binary components do not undergo this dynamical mass 
transfer.

\section{Evolutionary tracks and comparison to detailed simulations}

\begin{figure}[t]
%\epsscale{.90}
%\plotone{bseron1}
\includegraphics[width=84mm]{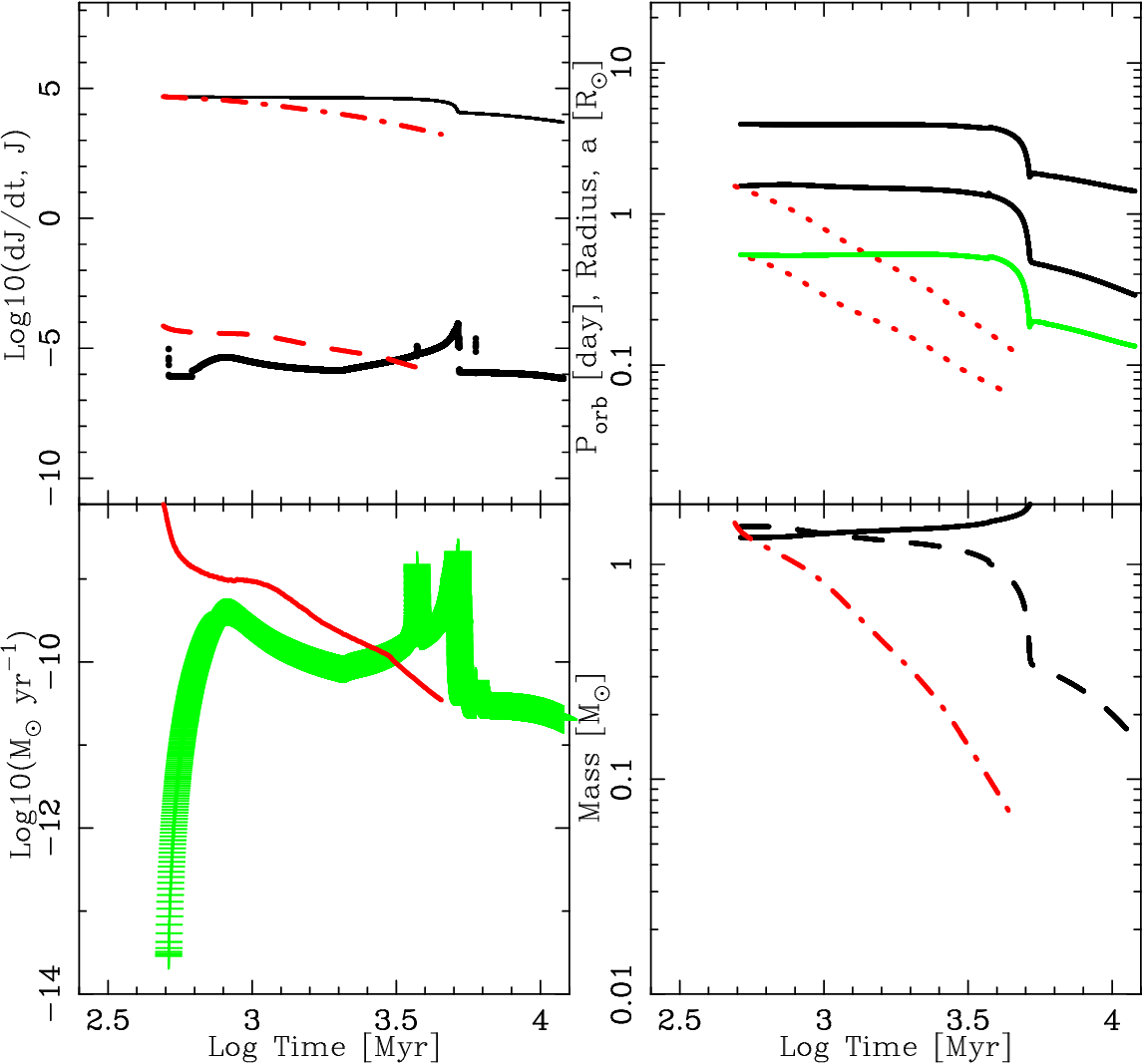}
\caption{Comparison between our rapid binary evolution code, BSE, and the detailed stellar evolution 
code with the inclusion of binary and pulsar evolution.  The system begins with $M_1=11.83~\msun$, 
$M_2=1.64~\msun$, $P_{\rm orb}=720~{\rm days}$ and zero eccentricity.  The mass transfer rate is 
that given in Hurley, Tout \& Pols (2002).
The top left panel depicts the evolution of the total 
change of angular momentum with time (black points) in units of $M_\odot R_\odot^2 {\rm yr}^{-2}$ 
in comparison with that used in the detailed evolutionary model (red dashed line).  
In units of $M_\odot R_\odot^2 {\rm yr}^{-1}$ the orbital angular momentum of both the 
model presented here (full thin line) and the detailed model (red dash-dot line) are illustrated.  
In the lower left panel, the mass accretion rate onto the NS (black points, all lie under green pluses) and 
loss rates from ablation (red points--which are close together) and the companion owing to `typical' mass transfer 
(green pluses) are illustrated as a function of time.  
The full red line traces the mass transfer rate from the detailed model.
The upper right panel illustrates the orbital separation (top black line) and the companion radius (lower black 
line) as a function of time.  Included is the orbital period in days of our system (green full line) 
and for comparison the orbital period and radius of the companion from the detailed model (lower and upper 
dashed red lines respectively).  
The lower right panel illustrates the variation of the NS mass (solid line) and the companion mass (dashed line) 
of our model as a function of time compared to the detailed model depicted by the red dash-dot line.
}
\label{f:bsevsron1}
\end{figure}

\begin{figure}[t]
%\epsscale{.90}
%\plotone{bseron2}
\includegraphics[width=84mm]{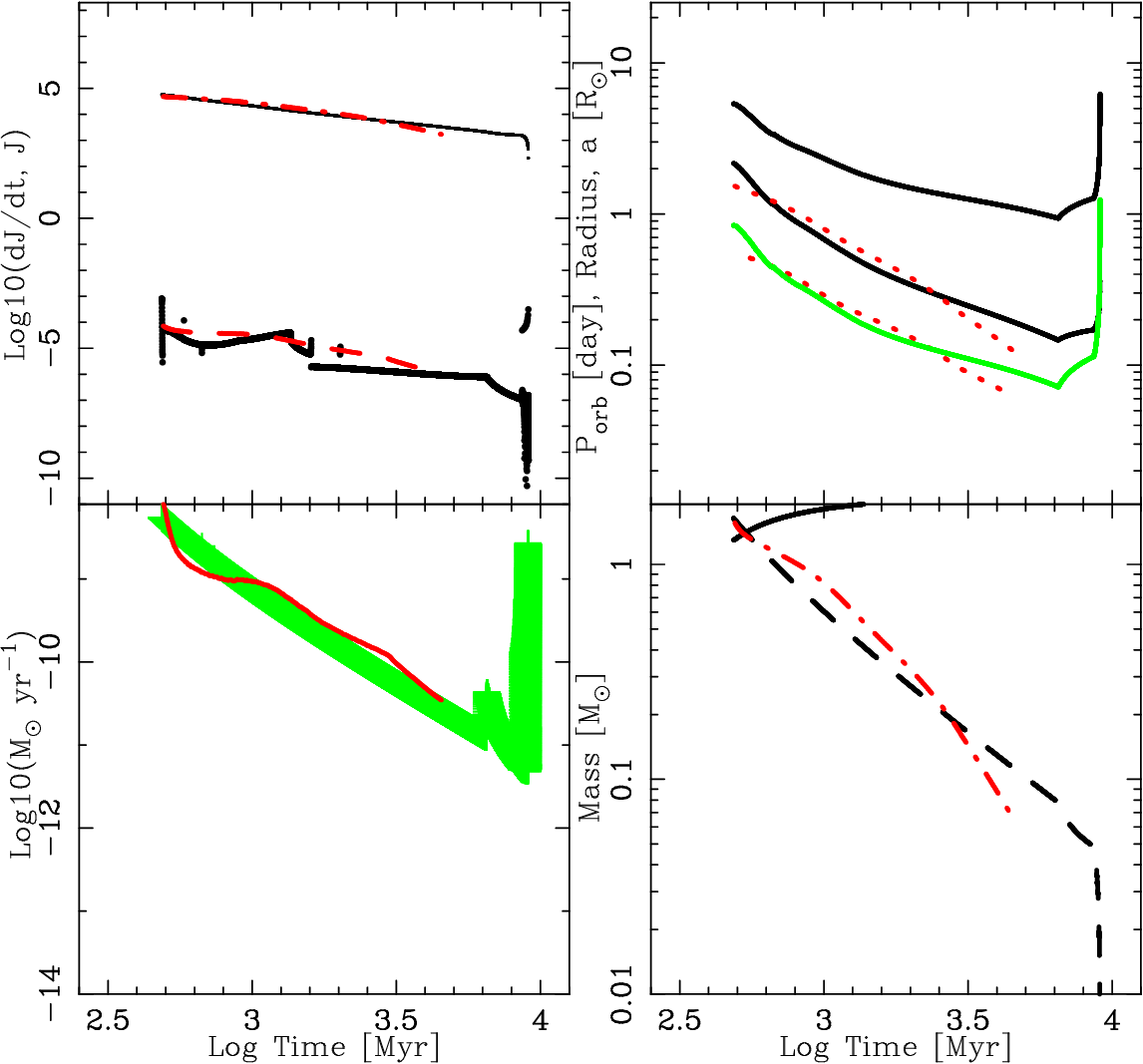}
\caption{
Comparison between our rapid binary evolution code, BSE, and the detailed stellar evolution code
with the inclusion of binary and pulsar evolution.
See caption of Fig.~\ref{f:bsevsron1} for details, however,
the mass transfer rate is updated from Hurley, Tout \& Pols (2002) equation and is now equation~\ref{e:newmdot}.
}
\label{f:bsevsron2}
\end{figure}

\begin{figure}[t]
%\epsscale{.90}
%\plotone{bseron3}
%\includegraphics{bseron3}
\includegraphics[width=84mm]{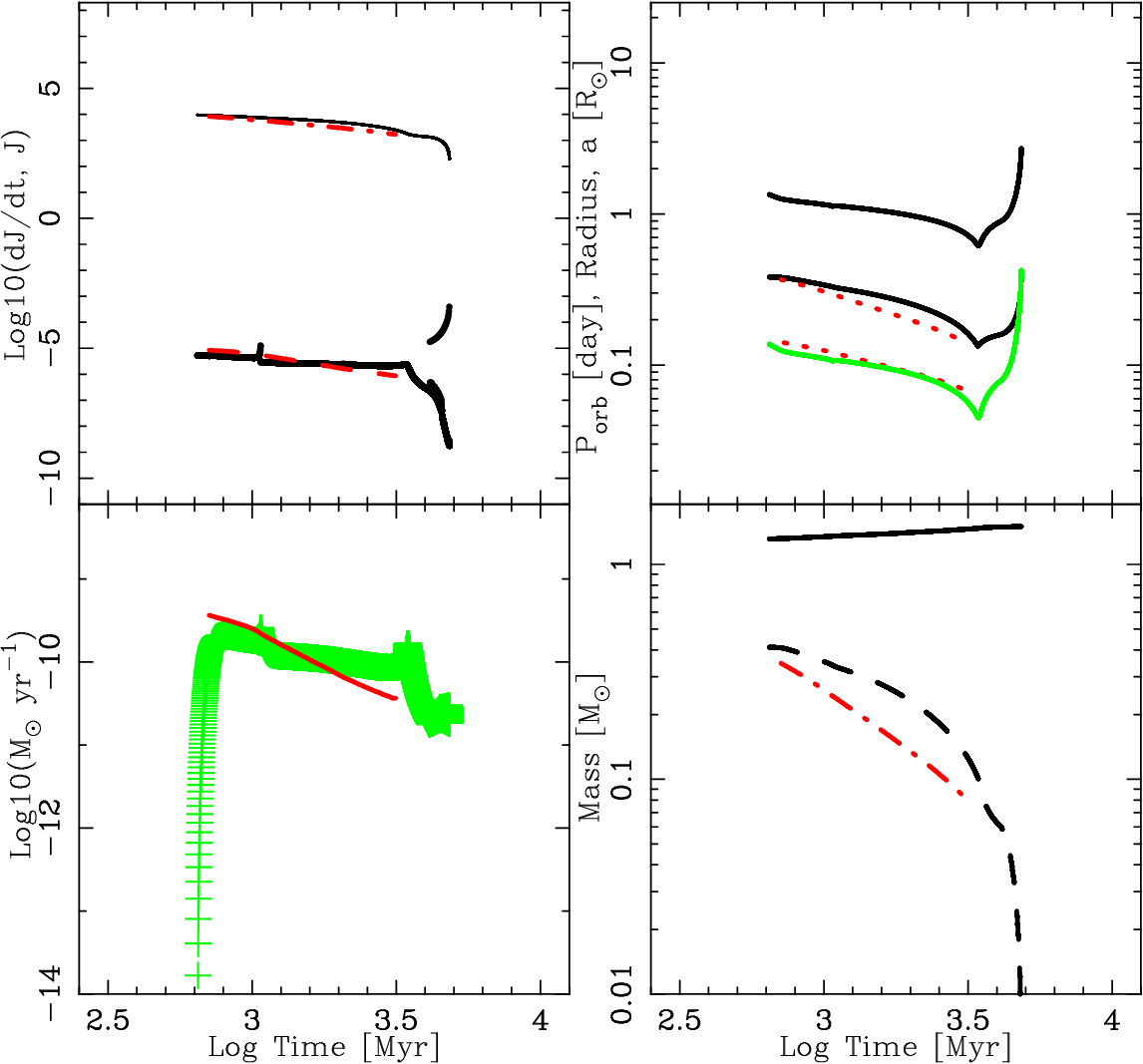}
\caption{Comparison between our rapid binary evolution code, BSE, and the detailed stellar evolution code
with the inclusion of binary and pulsar evolution.
The system begins with $M_1=8.936~\msun$, $M_2=0.4~\msun$, $P_{\rm orb}=842~{\rm days}$ and zero eccentricity.
The mass transfer rate is that given in Hurley, Tout \& Pols (2002).
The top left panel depicts the evolution of the total 
change of angular momentum with time (solid black points) in units of $M_\odot R_\odot^2 {\rm yr}^{-2}$ 
in comparison with that used in the detailed evolutionary model (red dashed line).  In units of $M_\odot 
R_\odot^2 {\rm yr}^{-1}$ the orbital angular momentum of both the model presented here (full thin line) and
the detailed model (red dash-dot line) are illustrated.  In the lower left panel, the mass accretion rate 
onto the NS (black points, all lie under green pluses) and loss rates from ablation 
(red points---which are close together) and the 
companion owing to `typical' mass transfer (green pluses) are illustrated as a function of time.  
The full red line traces the mass transfer rate from the detailed model.
The upper right panel illustrates the orbital separation (top black line) and the companion radius (lower black 
line) as a function of time.  Included is the orbital period in days of our system (green full line) 
and for comparison the orbital period and radius of the companion from the detailed model (lower and upper 
dashed red lines respectively).  
The lower right panel illustrates the variation of the NS mass (solid line) and the companion mass (dashed line) 
of our model as a function of time compared to the detailed model depicted by the red dash-dot line.
}
\label{f:bsevsron3}
\end{figure}

\begin{figure}[t]
%\epsscale{.90}
%\plotone{bseron4}
%\includegraphics{bseron4}
\includegraphics[width=84mm]{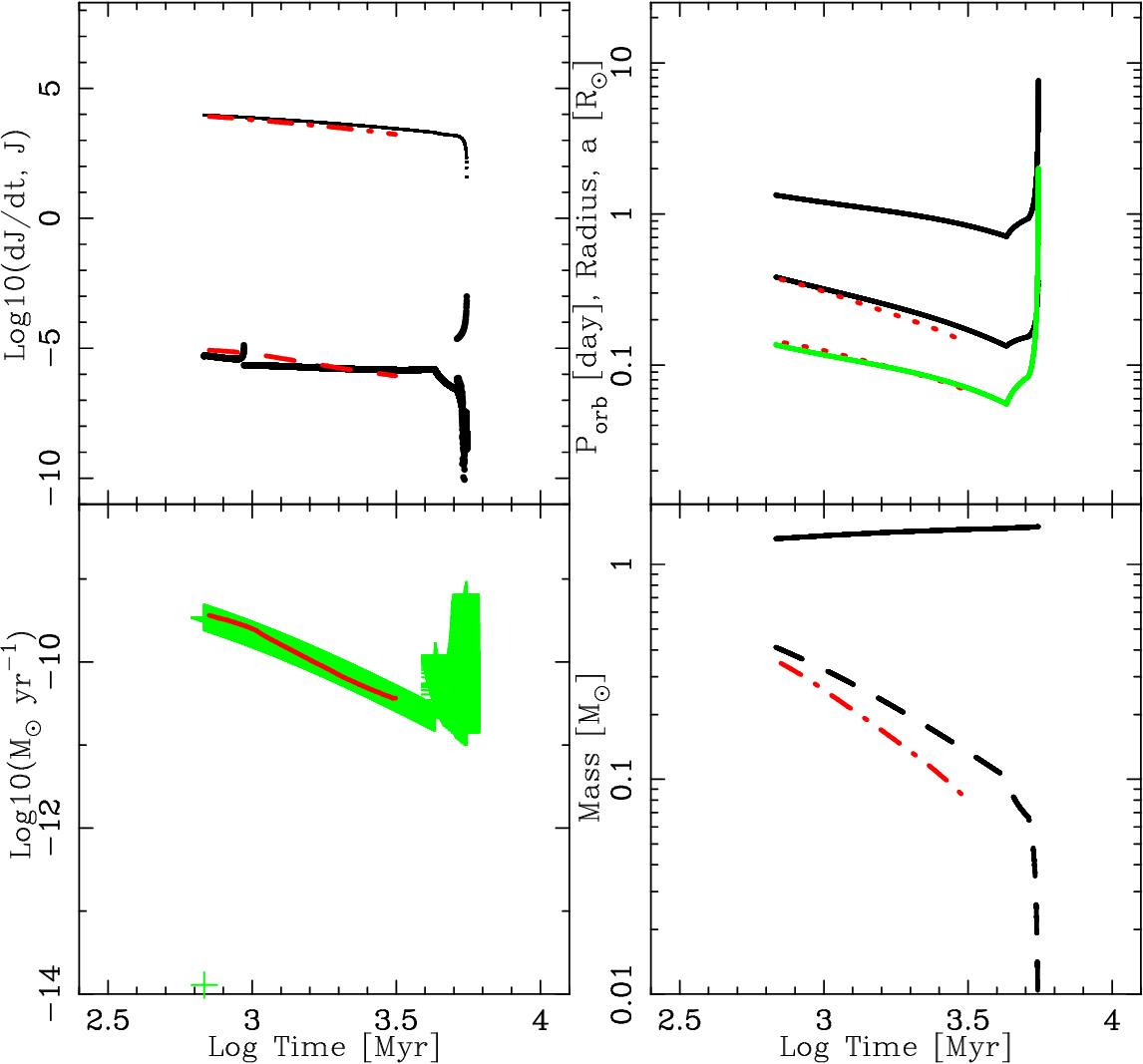}
\caption{
Comparison between our rapid binary evolution code, BSE, and the detailed stellar evolution code
with the inclusion of binary and pulsar evolution.
See caption of Fig.~\ref{f:bsevsron3} for details, however, the mass transfer rate is updated from 
Hurley, Tout \& Pols (2002) equation and is now equation~\ref{e:newmdot}.
}
\label{f:bsevsron4}
\end{figure}

\begin{figure}[t]
%\epsscale{.90}
%%\plotone{Fig5}
%\plotone{bseLin}
%\includegraphics{bseLin}
\includegraphics[width=84mm]{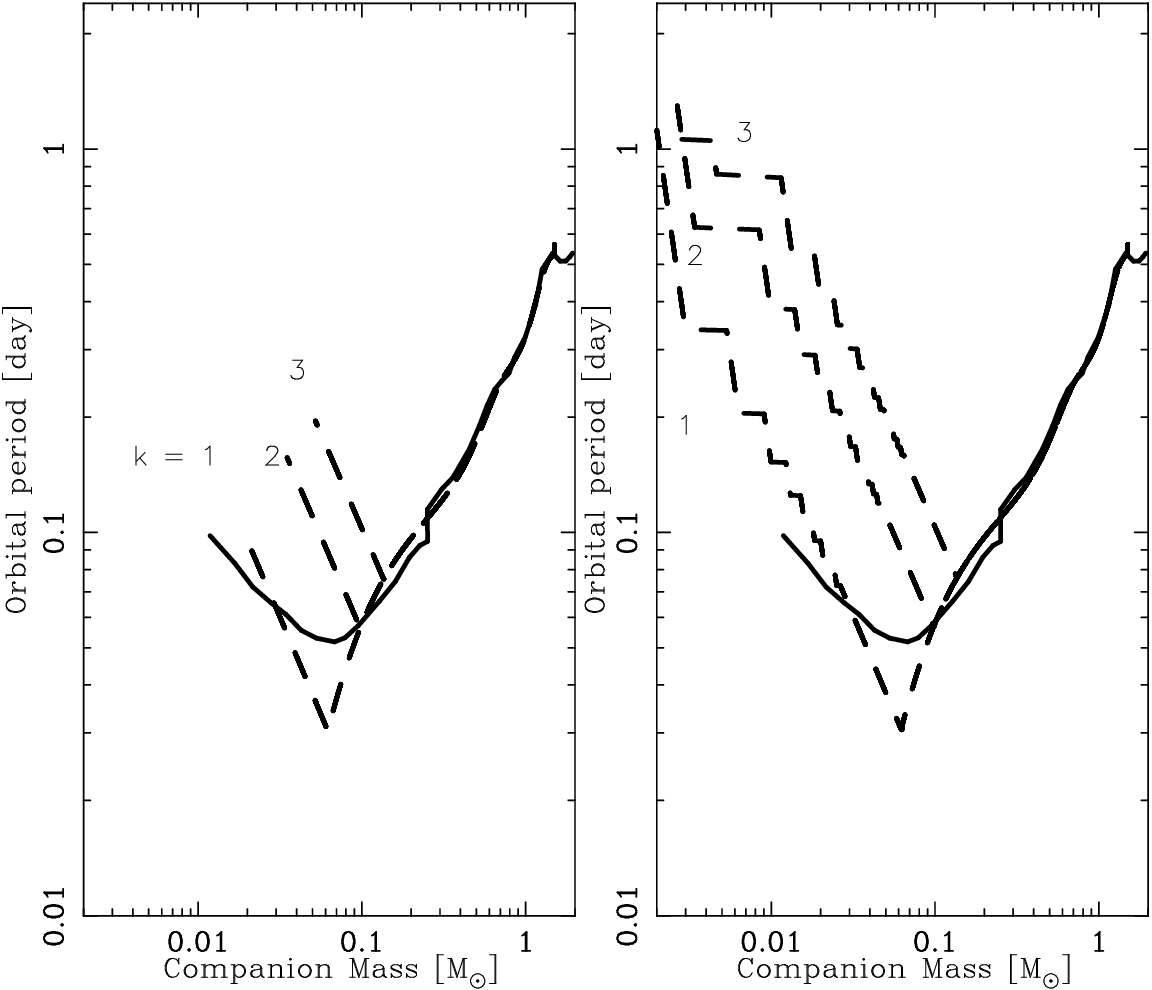}
\caption{
The evolution from the formation of the NS for the binary evolution based on 
the population synthesis code and the results from Lin et al. (2011).
Direct comparisons in the orbital period versus companion mass are illustrated in both panels.  
The solid line corresponds to an evolution with a $2\msun$ companion in Lin et al. (2011) as depicted 
in their fig.~2.  
The left panel includes models from the population synthesis without ablation or propeller 
evolution (see dashed lines for $k=1,~2~{\rm and}~3$).  
The right panel corresponds to similar models, but 
with the ablation and propeller mechanism included assuming $k=1$, 2 and 3. 
}
\label{f:Fig1.4}
\end{figure}

To determine the adequacy of the simplified prescription used for the binary evolution in the 
population synthesis, we compare our results to those obtained from a binary evolution code used 
in McDermott \& Taam (1989) and with results presented in Lin et al. (2011) in 
Figs.~\ref{f:bsevsron1}-\ref{f:bsevsron4} and Fig.~\ref{f:Fig1.4} respectively.  For comparison to 
McDermott \& Taam (1989) we examine two systems with companion masses at either end of the low 
mass X-ray binary companion mass range.  Figs.~\ref{f:bsevsron1} and \ref{f:bsevsron2} has a 
companion mass of $M_2 = 1.64\msun$ when the long lived mass transfer phase begins, while 
Figs.~\ref{f:bsevsron3} and \ref{f:bsevsron4} have $M_2 = 0.4\msun$ when mass transfer is first 
initiated.
The evolution is shown from the onset of the LMXB phase with $k=2$, where the pulsar spin period 
and magnetic field are $20~$s and $1.1\times10^{12}~$G respectively.  The detailed evolution of the 
donor was only followed to a mass $\sim 0.07 \msun$ because of the numerical issues in the equation of state. 
The magnetic braking orbital angular momentum loss methods are the same for both codes following 
Eqn.~\ref{e:mb}.  It is important to note that in BSE the steady nuclear mass transfer rate is set 
by a simple function of the companion mass and ratio of companion radius to its Roche lobe, and 
for compact accretors there is an additional term dependent on the accretor's radius given by 
(see Hurley, Tout \& Pols 2002; Eqn. 58), 
\begin{eqnarray}
\dot{M} = \frac{3\times10^{-3}}{max\left( R_1,10^{-4}\right)}&&[min\left( M_2,5.0 \right)]^2 \nonumber \\
&&[ln\left( R_2/R_{\rm 2RL}\right)]^3\msun~{\rm yr}^{-1}.
\label{e:bsemdot}
\end{eqnarray}
 
Within Fig.~\ref{f:bsevsron1} we artificially set the separation after the NS explosion such that 
mass transfer begins so that the MS star is approximately $500~$Myrs old in both models.  We see 
that the two models clearly diverge from the initial time step onwards.  It is the mass transfer 
rate that drives these divergent evolution solutions of this system, and it is here that the two 
calculations differ greatly.  To compensate for this difference we modified the mass transfer 
rate on nuclear timescales within BSE for NS and BH accretors by,
\begin{equation}
\dot{M}_{\rm new} = \frac{2000}{3} \dot{M}
\label{e:newmdot}
\end{equation}

This alternate evolution is shown in Fig.~\ref{f:bsevsron2} where we also ensure that the system 
remains in RLOF if the companion radius is decreasing with decreasing mass.  As a consequence we 
allow mass transfer during the time in which magnetic braking ceases to act as an angular momentum 
sink.  However, as can be seen in Figs.~\ref{f:bsevsron1} and \ref{f:bsevsron3} the original BSE 
method still allowed mass transfer during this time as well. As such we allow the magnetic 
braking process to occur continuously during the detailed calculations shown here.  A better 
treatment of this phase will be pursued at a later date.

The time evolution of the system in Fig.~\ref{f:bsevsron2} is now similar when comparing the 
two codes. Although there is some discrepancy in the mass transfer rates over time, we 
feel that BSE adequately captures the long-term evolution of the mass transfer rate.
The difference in radius and, therefore, separation arises because of the mass difference  
at each point in time.  

It is interesting to note that the pulsar does, eventually, begin to ablate its companion once it 
becomes a partially degenerate hydrogen-rich star. At this point the mass loss of the companion 
rapidly increases and the change in angular momentum begins to alternate between high and low values 
as the binary system attempts to adjust to this new phase of mass loss.

Fig.~\ref{f:bsevsron3} depicts the temporal evolution of our $0.4\msun$ companion mass system with the 
original mass transfer scenario of BSE.  We artificially set the separation after the NS explosion 
such that mass transfer begins so that the MS star is approximately $630~$Myrs old.  In this case, 
the mass transfer rate does quite well in evolving NS-LMXBs, when compared with the detailed 
evolution code. However, there are differences in the mass transfer rate at early and late times and 
as such we again adjust the mass transfer rate and algorithm.
We note that the first small drop in mass transfer rate of the BSE model (lower left corner 
of Fig.~\ref{f:bsevsron3}) is where magnetic braking terminates.

Increasing the mass transfer rate in BSE (Eqn.~\ref{e:newmdot}) again ensures a closer fit over time 
as compared to our detailed model.  The slight divergence in the companion mass is driven by slight 
differences in mass transfer, along with the alternate mass required for the systems to begin with 
the same orbital separation.  Better agreement would require fine tuning of BSE, however, 
it will not adversely affect our conclusions or rates as the time evolution of the ablation 
phase should not be overly affected.  Instead, it simply affects the time at which ablation occurs.
Because our population synthesis calculations have a flat distribution of birth ages, changes in 
the onset of ablation should not alter our statistics.

We have also compared the results of our binary evolution model with the recent study of Lin et al 
(2011).  Although there is insufficient information provided to make direct comparisons as in Fig. 4, 
we can compare the evolution in terms of the orbital period versus companion mass.  In 
Fig.~\ref{f:Fig1.4}, we take our initial system characterized by an $11.83\msun$ primary and a 
$1.5\msun$ secondary in a $740~$day orbit.  Evolution through a common envelope and supernova 
explosion leads to the formation of a system in which a $\sim 1.33\msun$ NS and $1.5\msun$ MS 
enters into a RLOF phase at a time of $500~{\rm Myr}$.  The pulsar is born with spin period of 
$0.3~$s and magnetic field of $2.7\times10^{12}~$G and, at the time of mass transfer, the pulsar 
spin period and magnetic field are characterized by $21~{\rm s}$ and $1.6\times10^{12}~{\rm G}$ 
respectively.  It takes only $\sim 1~$Myr for the pulsar to come into spin equilibrium.
In comparison to the Lin et al. (2011) model it can be seen in both panels of Fig.~\ref{f:Fig1.4} 
that the evolution is very similar during the phase corresponding to the decrease of orbital period 
with decreasing companion mass.  The major difference is seen in the phase corresponding to the 
increase in the orbital period with decreasing mass, with large values of $k$ departing from the 
detailed calculations of Lin et al. (2011), consistent with earlier work by King et al. (2005), but 
also places the models closer to where many black widow pulsars are observed.  

Taking into account the effects of propeller and ablation leads to lower masses and correspondingly 
longer orbital periods as shown in the right panel of Fig.~\ref{f:Fig1.4}.  However, the orbital 
expansion is governed by the mass-radius relationship, and therefore the value of $k$, rather than 
the angular momentum losses from propeller evolution and ablation of the companion.  It is 
interesting to note that the case for $k=1$ in the lower right panel leads to an evolution which lies 
in the vicinity of the Lin et al (2011) curve but extends the companion masses evolution to lower 
masses and longer orbital periods.  

\subsection{Example evolutionary tracks of black widow phase}
\label{s:evex}

Prior to our analysis of the population synthesis characteristics of black widow systems, we
first determine the companion mass-orbital period parameter space where the systems are expected
to form, building on the earlier work of King et al. (2005).
Here, we take the equations defined above that describe the transition to the ablation phase and 
with the simple assumptions for the mass-radius relation and mass transfer rate, calculate the 
critical transition points.
  
We assume the companion fills its Roche lobe with its radius determined from the companion 
mass-radius relationship of
\begin{equation}
R_{\rm MS} = 0.0128 \left(1+X \right)^{5/3} k \left( \frac{M_{\rm d}}{M_\odot} \right)^{-1/3}\rsun.
\end{equation}
This assumption leads to an estimate of the mass transfer rate, assuming the change in angular 
momentum is driven by gravitational radiation, such that 
$\dot{M}_{\rm gr2}/M_2 = \dot{J}/\left[ J(2/3-M_2/M_{\rm NS} \right]$.
This rate is then compared to the critical disk stability limit (Eqn.~\ref{e:lasota2}) of 
Lasota et al. (2008) and to the gamma-radiation equation (Eqn.~\ref{e:gammamdot}) of 
Takata, Cheng \& Taam (2010).  The transition region is determined by $\dot{M}_{\rm gr2}$ lying 
below the values of both the critical limits.
We assume a NS mass of $1.4\msun$, eccentricity of zero, pulsar spin period of $5~$ms, and 
pulsar magnetic field of $1\times10^8~$G.  This standard setup is shown with the three lines for 
various companion masses and bloating factors.  In addition, we also vary the NS mass, pulsar 
spin period and magnetic field.

\begin{figure}[t]
%\epsscale{.50}
%\plotone{cmd}
%\includegraphics{cmd}
\includegraphics[width=84mm]{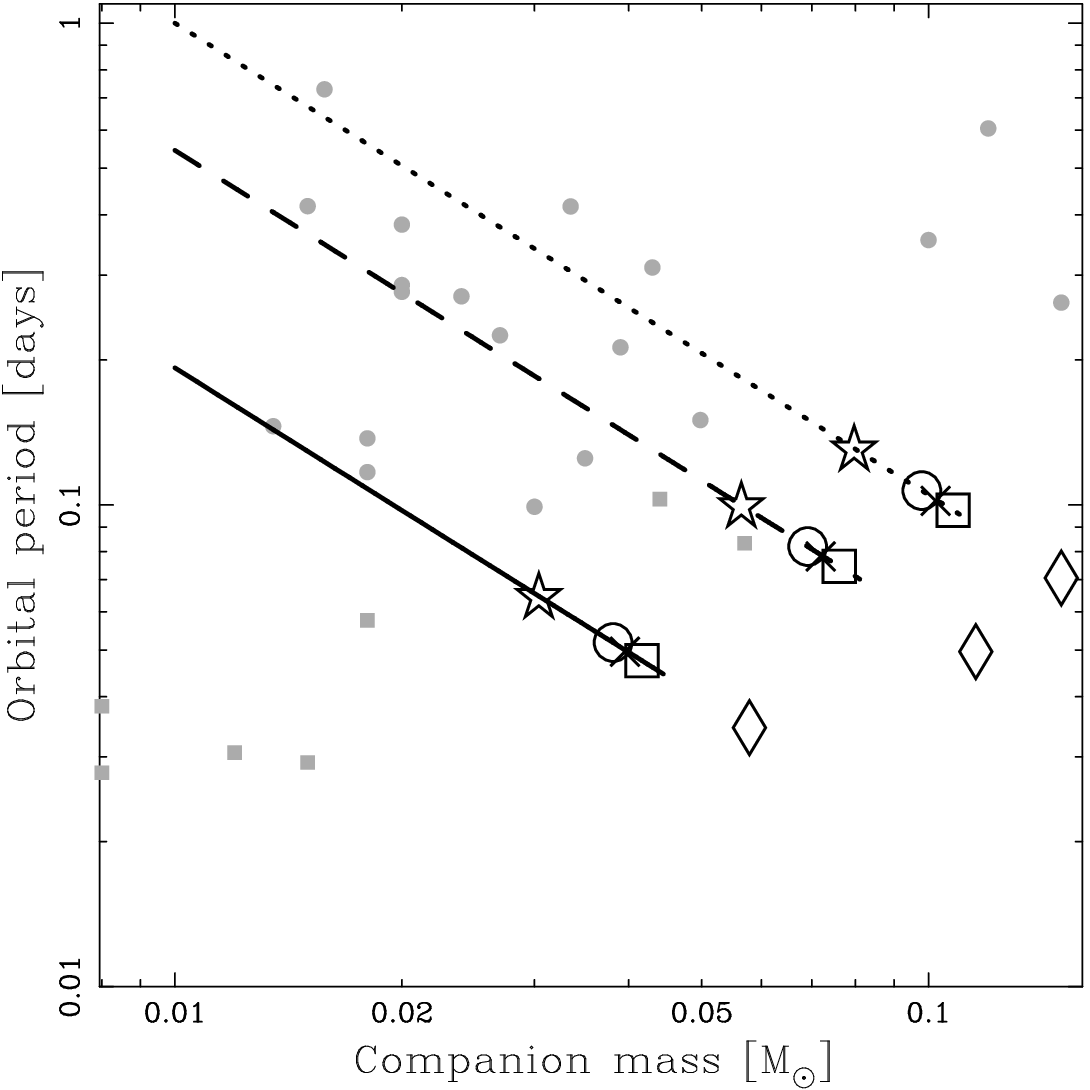}
\caption{
Typical estimated values of companion mass and orbital period
when ablation begins.
The sensitivity to model assumptions in determining the transition phase are also depicted.
A NS mass of $1.4\msun$, eccentricity of zero, pulsar spin period of $5~$ms,
pulsar magnetic field of $1\times10^8~$G are adopted.
The standard setup is shown with the solid, dashed and dotted lines corresponding 
to $k=1$, $2$ and $3$ respectively.
Results are also shown for a NS mass to $2\msun$ (cross), pulsar spin period to $1~$ms (circle),
pulsar magnetic field of $5\times10^7~$G (square), magnetic field of $10^{10}~$G (diamond),
and a combination of these changes with NS mass 
$2\msun$, spin period of $1~$ms and field of $5\times10^7~$G (star).
Each of these assumptions are calculated for each bloating factor.
Underlaid beneath the theoretical calculations in grey are observations, with circles representing 
rotation powered MSPs taken from Table~\ref{t:tableobs} and the ATNF (Manchester et al. 2005) 
catalogue while squares depict known accretion powered MSPs from the Ritter \& Kolb (2003) catalogue. 
Note that the three highest companion mass systems are not black widow systems but detached MSP-WDs,
where the MSP was most likely spun-up by a giant or subgiant star.
}
\label{f:critmd}
\end{figure}

Upon examining the trends exhibited in Fig.~\ref{f:critmd}, it is evident that the onset of the 
ablation phase is most sensitive to the degree of thermal bloating of the companion.  Lower values 
of the bloating result in shorter orbital periods and lower companion masses at the transition. 
A higher magnetic field strength leads to an increase of the critical mass transfer rate that 
determines when a pulsar may begin to halt accretion (Eqn.~\ref{e:gammamdot}).
It is this equation that, on average, determines when ablation will occur because it results in a 
lower critical mass transfer rate than that given by the disk stability limit (if one 
assumes 'typical' parameter values).  Similarly, decreasing the magnetic field and spin period 
increases the orbital period at which the change in system state occurs.  The transition region 
depends little on the NS mass.

\begin{table*}
\small
% \centering
% \scriptsize
% \begin{minipage}{140mm}
  \caption{Characteristics of the observed black widow pulsar systems ordered in ascending orbital period.
  \label{t:tableobs}}
%  \begin{tabular}{crrrrrrrr}
  \begin{tabular}{@{}crrrrr@{}}
%  \hline
\tableline
Name & $M_2~[\msun]$ & porb~[day] & P~[s] & $\dot{P}$~[s/s] & Ref.\\
\hline 
%J1719-1438 & 0.0013 & 0.091 & 0.0058 & 7.3e-21 & Bailes et al. (2010) \\% # J1719-1438 Bailes: diamond planet pulsar.
J2051-0827 & 0.03 & 0.099 & 0.00451 & 1.27e-20 & Stappers et al. (1996)\\% #J2051-0827 --Old BW
J1544+4937 & 0.018 & 0.117 & 0.00216 & --- & Kerr et al. (2012), Roberts (2012) \\% J1544+4937 F (Bhattacharaya et al. 2012)
J2047+10F & 0.035 & 0.125 & 0.00429 & --- & Ray et al. (2012), Roberts (2012) \\% J2047+10 F Ray et al. (2012)
J0023+09F & 0.018 & 0.138 & 0.00305 & --- & Hessels et al. (2011), Roberts (2011) \\% (J0023+09F)
J2241-5236 & 0.014 & 0.146 & 0.00219 & 6.64e-21 & Keith et al. (2011) \\% J2241-5236 (from Fermi, south)
J1810+17F & 0.050 & 0.15 & 0.0023 & --- & Hessels et al. 2011, (Roberts 2011) \\% (J1810+17F)
J2256-1024F & 0.039 & 0.213 & 0.00229 & --- & Boyles et al. (2011), Gentile (2012) \\% (J2256-1024F)
J1124-3653 & 0.027 & 0.225 & 0.00241 & --- & Hessels et al. (2011), Roberts (2012) \\% J1124-3653 F (Hessels et al. 2011).
J1301+0833F & 0.024 & 0.271 & 0.00184 & --- & Ray et al. (2012), Roberts (2012) \\% J1301+0833 F (Ray et al. 2012).
J1446-4701 & 0.02 & 0.277 & 0.00219 & --- & Keith et al. (2012) \\%#1446-4701 (Keith et al. 2012)
J0610-2100 & 0.02 & 0.286 & 0.00386 & 1.24e-20 & Burgay et al. (2006)\\%#J0610-2100 --Old BW
J1731-1847 & 0.043 & 0.311 & 0.00234 & 2.49e-20 & Bate et al. (2011) \\% J1731-1847
J1959+2048 & 0.02 & 0.382 & 0.00161 & 1.69d-20 & Fruchter et al. (1988)\\%# B1957+20 (J1959+2048) [This is B1957+20(F) according to Roberts 2012] --Old BW
J2214+3000 & 0.03 & 0.417 & 0.00312 & 1.401e-20 & Ransom et al. (2011) \\% (pdot = dv/dt/(v^2); assumes Mns = 1.4Msun; not 1.35Msun like above!!!; PSR J2214+3000)
J2234+0944F & 0.015 & 0.417 & 0.00363 & --- & Roberts (2012) \\% J2234+0944 F Keither et al. (2012a)
J1745+1017F & 0.016 & 0.729 & 0.00265 & --- & Barr et al. (2013), Roberts (2011) \\% (J1745+10F)
%\hline
\tableline
\end{tabular}
\end{table*}

Uniting together into one example each of the effects discussed above (and assuming a magnetic 
field of $5\times10^{7}$G) increases the orbital period of where the transition region occurs and places 
it between the grey squares and circles.
The observed sources are listed in Table~\ref{t:tableobs}.
As seen from Eqns.~\ref{e:gammamdot} and \ref{e:lasota2} the parameters that lead to the largest variation in 
transition region are associated with the NS radius and accretion disk radius.

\begin{figure}[t]
%\epsscale{.90}
%\plotone{4plttest}
%\includegraphics{4plttest}
\includegraphics[width=84mm]{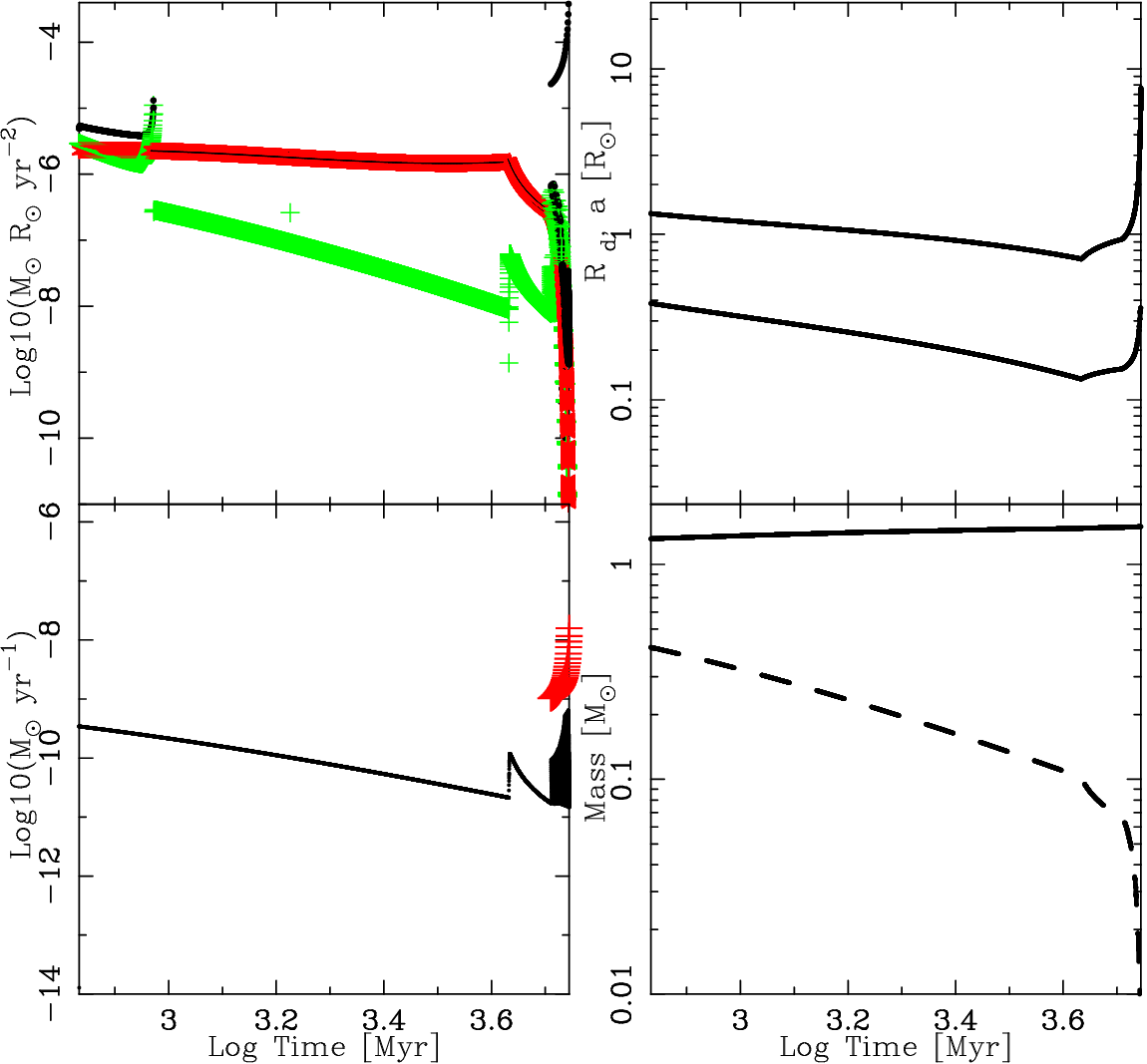}
\caption{Detailed evolution of a system initially characterized by $M_1=8.936~\msun$, $M_2=0.4~\msun$, 
$P_{\rm orb}=842~{\rm days}$ and zero eccentricity.
The evolution is shown from the onset of RLOF, where we assume that $k=2$, $f_{\rm e}=0.001$ ($0.1\%$) 
and $f_{\rm eq}=0.5$.  
The top left panel depicts the evolution of the total change of angular momentum with time (black points) and, 
for comparison, the change in orbital angular momentum owing to tides (green pluses) and gravitational 
radiation (red asterisk) with time. 
In the lower left panel, the mass accretion rate onto the NS (black points) and loss rates from 
ablation (red plus) are shown.  The top right panel illustrates the orbital separation (top line) and the 
companion radius (lower line) as a function of time. The lower right panel illustrates the variation of the NS 
mass (solid line) and the companion mass (dashed line) as a function of time.}
\label{f:Fig1.1}
\end{figure}

\begin{figure}[t]
%\epsscale{.90}
%\plotone{4plttestMstar}
%\includegraphics{4plttestMstar}
\includegraphics[width=84mm]{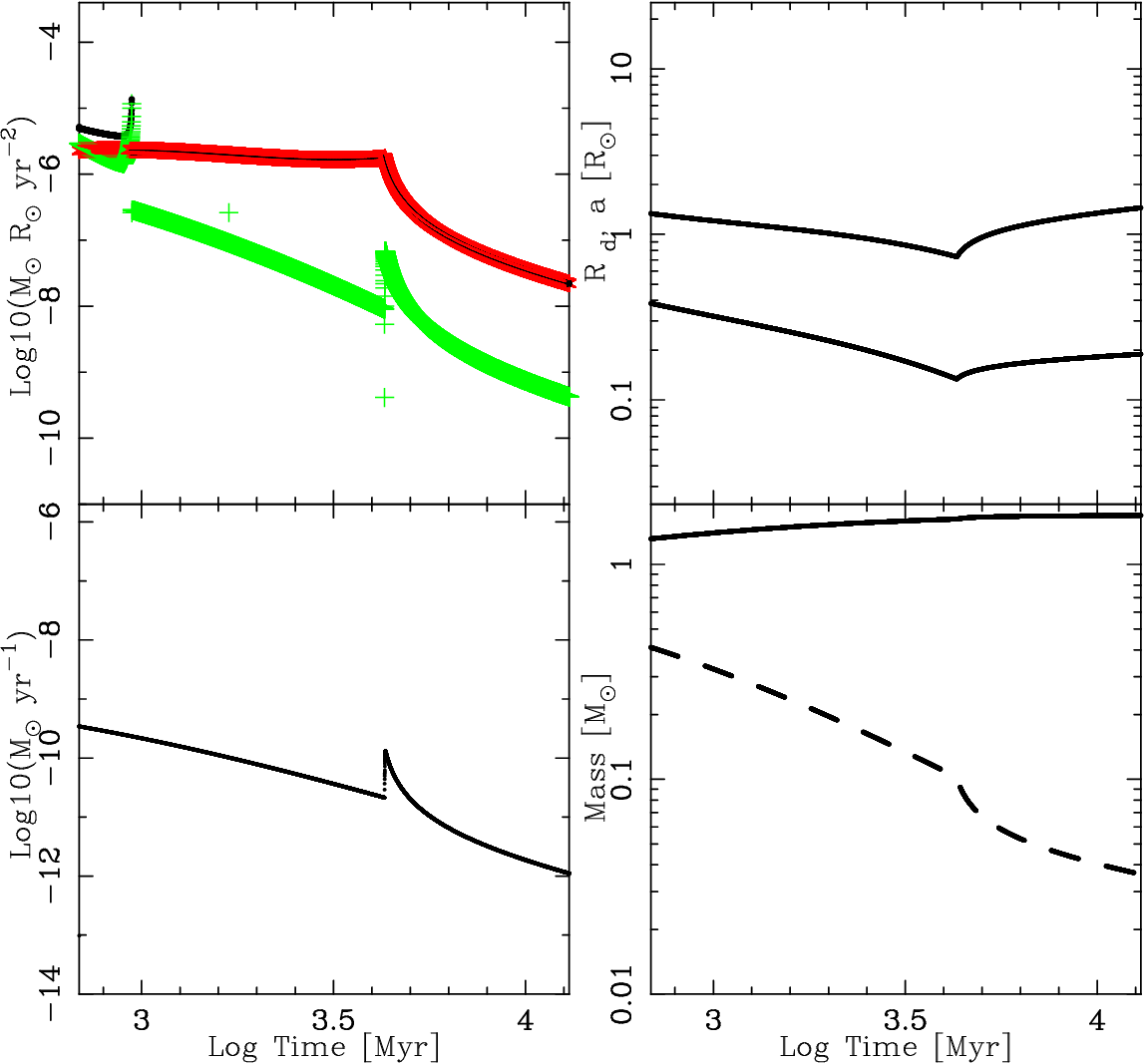}
\caption{Except for $M_\star = 10^{-6}$, all else follows Fig.~\ref{f:Fig1.1}.}
\label{f:Fig1.2}
\end{figure}

To examine the sensitivity of the evolutionary pathways for particular assumptions of 
our population synthesis modeling we vary parameters of our model and provide details 
for the variation of $R_{\rm d}$, $a$, $\dot{M}$ and $\dot{J}$ with respect to time in 
Figs.~\ref{f:Fig1.1} and \ref{f:Fig1.2}.  
As an example, we use the system depicted in Fig.~\ref{f:bsevsron4} (see caption for details).
The NS within this simulation is born with $P_{\rm s} = 0.3~{\rm s}$ and 
$B_{\rm s} = 2.7\times10^{12}~{\rm G}$ and formed soon after the system emerged from 
the common envelope phase with an orbital period of $0.2~{\rm day}$.  
At the start of mass transfer the NS has $P_{\rm s} = 24~{\rm s}$ and 
$B_{\rm s} = 1.3\times10^{12}~{\rm G}$ and at the simulations end
it has $P_{\rm s} = 0.0007~{\rm s}$ and $B_{\rm s} = 9.8\times10^{7}~{\rm G}$.
In this case, the system evolves to the semi-degenerate hydrogen-rich track and is analogous to
a cataclysmic variable system, but with a NS compact object. 
The system evolves to the black widow phase, and the pulsar fully ablates its companion.
The system does not evolve by regularly passing through the ablation phase, but instead
undergoes alternating phases of equilibrium mass transfer and ablation.
We also point out that qualitatively, the mass accretion rate onto the NS as shown in 
Fig~\ref{f:Fig1.1} is similar in shape and magnitude to those calculated by Rappaport, Joss 
\& Webbink (1982), whose binary mass transfer was driven by gravitational radiation only.  

Upon further examination of Fig.~\ref{f:Fig1.1} a number of features are present in the panel 
illustrating the mass accretion rate as a function of time, which are associated with similar 
changes in the variation of angular momentum.  The total change in angular momentum with time 
(black points in Fig~\ref{f:Fig1.1}) closely follows the orbital decay owing to gravitational 
radiation (red stars), except for an unexpected peak as compared to the mass transfer results found
by Rappaport, Joss \& Webbink (1982). 
The sudden increase in the mass accretion rate reflects the slow
response of the orbit to the change in stellar structure of the companion 
becoming partially degenerate and leading to the change in sign of the radius mass derivative.
Because $\dot{M} \propto 
\left(\log{R_{\rm d}/R_{\rm RL~d}}\right)^3$, where $R_{\rm RL~d}$ is the donor's 
Roche lobe radius, the mass transfer rate increases.
This result follows from the fact that the Roche lobe radius is not forced to be the stellar 
radius exactly, when the radius increases.
Therefore, over a small number of time steps the orbit continues to decay while the 
companion radius increases.  
These features occur over a relatively short time scale and are due to our assumptions 
in the code and the time explicit method of time integration. 
It is unlikely that their influence is significant as the perturbations are small.  
For the most part the separation of the two stars follows the change in companion radius.  

We have also examined the sensitivity of our results to the magnetic field decay rates.
Specifically, lowering the value of $M_\star$ increases the rate at which the magnetic 
field decays with accreted mass.
The effect of setting $M_\star = 10^{-6}$ is shown in Fig.~\ref{f:Fig1.2} and results in 
an evolution where 
the pulsar never has the power to ablate its companion.
The mass transfer rate evolution is similar to that described in Fig.~\ref{f:Fig1.1}.

To examine the dependence of the evolution of the systems to the equilibrium spin relation, we 
increased $f_{\rm eq}$ from 0.5 to 1.  That is, the NS is on the equilibrium line at higher magnetic 
fields, or for the same field strength the equilibrium spin is at a faster rate.  
This causes the ablation process to occur at a later time (by $\sim400$Myr).  Interestingly, the 
time spent in ablation is shorter as more angular momentum losses lead to higher mass transfer rates, 
which in turn depletes the companion faster.  Ablation lasts for only a few tens of millions of years.

Adopting a lower efficiency of conversion of pulsar luminosity to mass loss by decreasing the value 
of $f_{\rm e}$ from 0.1\% to 0.01\%  decreases the fraction of mass ablated from the companion by 
up to an order of magnitude at early times, and reduces the rate of orbital angular momentum loss. 
As a consequence the system lives longer by $\sim 1~$Gyr.

If we decrease $k$, the orbital evolution of the companion, during ablation, is affected. 
In this case, ablation of the companion occurs prior to the orbital period minimum. 
Here, the orbital angular momentum losses drive this system to the period minimum within 
$100~$Myr from the onset of ablation.
This evolution was not explored in our previous examination of the transition phase 
into ablation and occurs at an orbital period of $0.05~$day and a companion mass of $0.098\msun$.
As an additional result associated with a decreasing value of $k$, we note that the mass 
ablation rate increases and the pulsar spends a greater time ablating its companion 
compared to the values in Fig~\ref{f:Fig1.1} because the two stars are now closer together.  In this 
case, the companion mass decreases more rapidly, quickly leading to the formation of an isolated MSP.

\section{Example evolutionary outcomes over a range of initial systems}

The results of our population synthesis simulations, and in particular two models which comprise 
our `standard' assumptions, are presented below.  These two models differ in one respect, viz., model 
A (B) does not (does) include propeller evolution and/or ablation of the companion.  For convenience, 
we provide the typical values for some of the more important parameters here.  Specifically, we 
evolve $N=3\times10^8$ binary systems.  For each CE we assume $\alpha = 3$, while we include a 
variable binding energy constant, $\lambda$, as discussed above.
NSs can be born via either core collapse or electron capture SNe (which includes accretion induced 
collapse), and we assume the resultant SN asymmetric kicks are drawn from a Maxwellian distribution 
(see \S 2.3). 
When a pulsar evolves to longer spin periods we allow the field to decay exponentially 
on a timescale of $\tau_{\rm B} = 900~$Myr, while in isolation and decay with mass 
accreted during the mass accretion phase with $M_\star = 1\times10^{-4}\msun$.
These models are based on the modified mass transfer rate provided in Eqn.~\ref{e:newmdot} 
for NSs accreting from MS stars.  We force this mass transfer to continue until a physical process 
such as the propeller phase leads to its termination.  Each system is born at a random time in 
the Galaxy's history, resulting in a flat distribution of birth ages between $0-12~$Gyr.

\begin{figure}[t]
%\epsscale{.90}
%\plotone{initp58m1m2}
%\includegraphics{initp58m1m2}
\includegraphics[width=84mm]{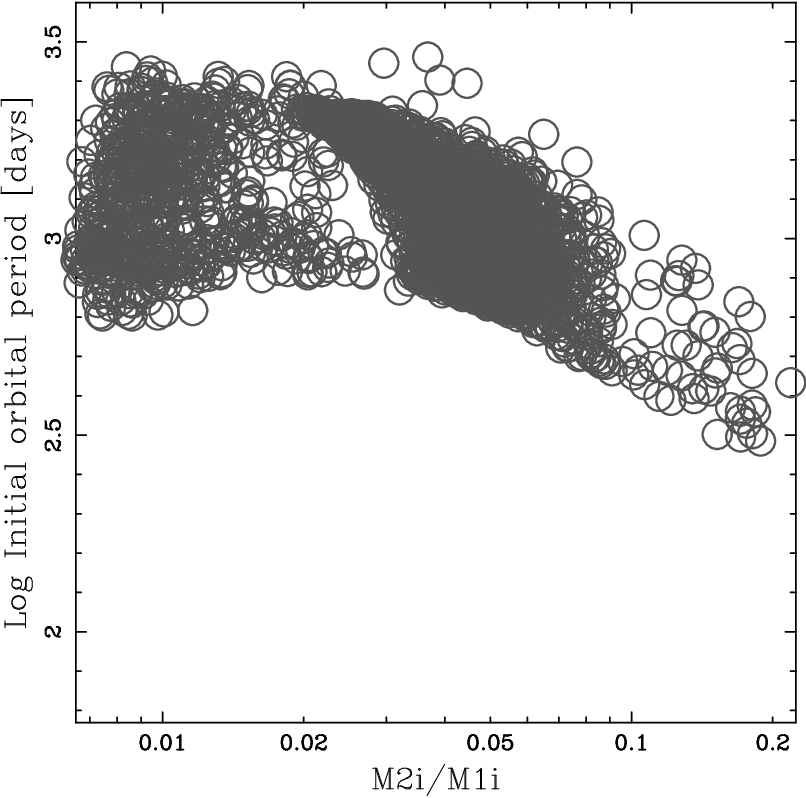}
\caption{Initial properties of mass ratio and log of the orbital period are shown for those systems
from Model B that are in a black widow system at the end of the simulation. 
}
\label{f:inittbq}
\end{figure}

Before examining the results from our population synthesis simulations we first briefly 
discuss the initial properties of those systems that become black widow MSPs in Fig.~\ref{f:inittbq}.
A number of interesting features of the initial systems required to form a black widow system are 
depicted. The most important and interesting feature gleaned from this figure is that small 
mass ratio systems are more prominent than higher ones in forming black widow systems.  The majority 
of progenitor NS masses reside between $8.5-10\msun$ with islands in the parameter space at $15\msun$ 
and $22\msun$. Most initial companions are less massive than $0.5\msun$.  This is primarily a 
selection effect, as it is easier for lower mass stars to reach binary properties (orbital 
periods and pulsar characteristics) that can initiate ablation of the companion in shorter periods 
of time than more massive companion stars. This results from the fact that lower mass 
companions must spiral in deeper in the envelope during the CE phase for successful ejection.

We find that there is no significant change to this finding when other parameters are modified.
The initial orbital periods of black widow systems mostly lie in a band of periods between 
$\sim650~$days and $\sim2500~$days, which does not change significantly with changes 
to $\alpha = 1$ and $\lambda = 0.5$. The only noticeable effect is a greater range of 
initial orbital period.  
A modification of the initial mass ratio distribution to a flat 
distribution leads to a reduction in the statistics.
However, the distribution of initial 
$P_{\rm orb}$ and $q$ still reflects that of Model B---with the highest density region lying 
at $q\sim0.04$, although there is a greater number of systems (relatively) residing at $q$ 
between $0.1-0.2$.  The initial orbital period distribution in this case is very similar to Model B.

To demonstrate the influence of mass loss from the system due to ejection from the accretion 
disk by propeller action and the ablation of the companion by the pulsar on the MSP evolution, we 
consider the numerical results from the population synthesis of the two models.  A binary MSP 
population corresponding to the standard mass transfer assumptions is adopted for model A, whereas 
model B includes propeller and ablation evolution.  We note that (i) model A is comparable to 
a typical model from the recent study of Hurley et al. (2010), but with our updated mass transfer 
rate and algorithm and (ii) there is a large orbital period dependence upon $k$.

\begin{figure}[t]
%\epsscale{.90}
%\plotone{pop58test2}
%\includegraphics{pop58test2}
\includegraphics[width=84mm]{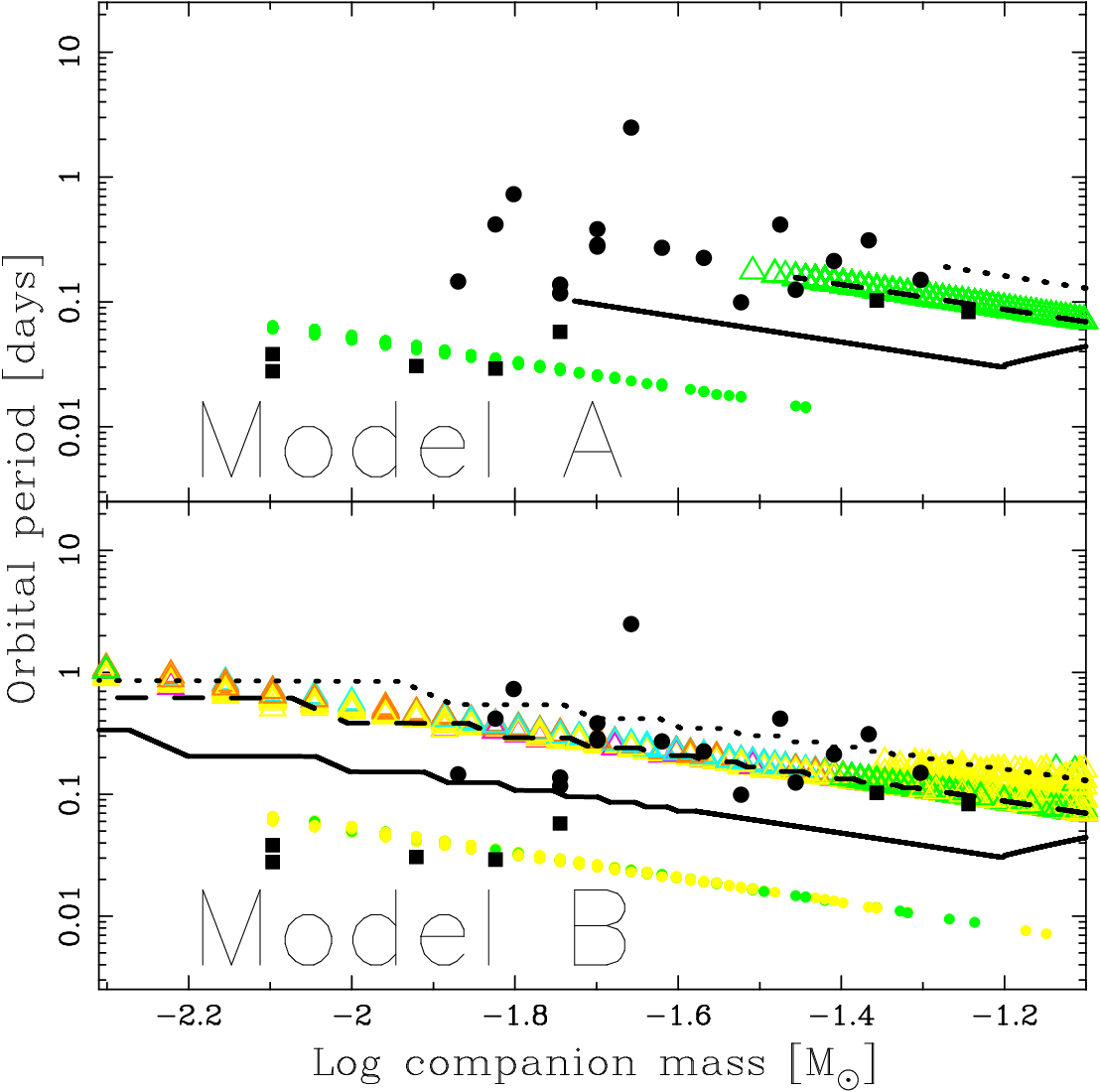}
\caption{Orbital period vs companion mass for models A (upper panel) and B (lower panel). Colored triangles 
represent low mass/semi-degenerate main sequence stars (we simply refer to these as MS-MSP systems), while 
colored dots represent He/CO/ONe WDs (which we refer to as WD-MSPs).  The different colors denote different 
phases of evolution; RLOF is green, mass transfer during spin equilibrium evolution is yellow, ablation while 
on the equilibrium line is magenta, ablation during propeller evolution is cyan and the detached phase is 
orange.  Over plotted on both panels are black points depicting observed systems.  
Full squares are X-ray binaries with known accretion powered MSPs taken from the 
Ritter \& Kolb (2003) catalogue.  
Full circles are known black widow pulsars taken from Table~\ref{t:tableobs} 
and making use of the ATNF catalogue (Manchester et al. 2005)
assuming ${\rm M}_{\rm NS} = 1.35$ and $i=60^\circ$.
The lines represent a particular system from the same model but with $k = 1$ (full line), $k=2$ 
(dashed line) and $k=3$ (dotted line).
The system begins with $M_1=8.936~\msun$, $M_2=0.4~\msun$, 
$P_{\rm orb}=842~{\rm days}$ and zero eccentricity.
}
\label{f:Fig1}
\end{figure}

To illustrate the importance of propeller and ablation evolution in the formation of rotation and 
accretion powered MSPs (pulsars with $P<0.03~{\rm s}$) we provide a comparison of models A and B 
in the orbital period/companion mass plane (see Fig.~\ref{f:Fig1}; where only 1/100 of the model 
statistics is shown for clarity).  Both models clearly exhibit the existence of two branches where 
the orbital period increases with decreasing companion mass.  The upper branch describes the MS-MSP 
population whereas the lower branch depicts WD-MSPs (UCXBs).  In model A all systems are mass 
transferring (i.e., LMXB's) and the separation of the two branches in orbital period results from 
differences in the donor radii (see \S 1).  In this model (where $k=2$), hydrogen-rich stars reach 
a mass of $\sim 0.025\msun$ within a Hubble time, which sets a lower mass limit to donors evolving 
from the MS branch.  We note that fully degenerate hydrogen-rich companions (where $k=1$) may 
reach masses $<0.01\msun$ within a Hubble time due to higher angular momentum losses associated with 
gravitational radiation at shorter orbital periods.  On the other hand, the results of model B reveal 
the existence of systems where the companion has detached from its Roche lobe and mass accretion 
has abruptly ceased.
In this model, radio pulsars form in systems extending to lower companion masses 
on the low mass semi-degenerate branch of the MS with the evolution driven by ablation of the donor 
stars while the systems are evolving on the equilibrium spin line.  We point out that the 
operation of the propeller mechanism is essential for facilitating the onset of pulsar activity and 
the subsequent ablation of the companion.

Many of the systems with partially degenerate hydrogen-rich companions in model B alternate 
between various evolutionary phases, however, the `accretion on the equilibrium' phases (yellow 
triangles) dominates the majority of times for most systems with the pulsar spending half of 
this phase `on' and in propeller mode so as to maintain equilibrium spin (see \S 2.1.2).
The influence of modifying the companion thermal bloating parameter, $k$, is illustrated in 
Fig.~\ref{f:Fig1}, where we show the result for $k=~$1, 2 and 3 on the binary population.  
As expected a smaller (larger) value of $k$ results in tighter (larger) orbits and lower (longer) 
minimum orbital periods.  We remark that empirically, and prior to any complicating effects 
associated with propeller or ablation, $k=2$ apparently matches the two observed X-ray binaries on 
this evolutionary branch (right most squares in Fig.~\ref{f:Fig1}).  The primary affect of $k$, in 
this context, is to cause enhanced expansion of the orbit with increasing values of $k$.  
Examination of Fig.~\ref{f:Fig1} and integration of Eqn.~\ref{e:ab_j} reveal that even with 
the operation of the propeller and ablation mechanism the orbital evolution is dominated by 
our choice of $k$.  Therefore, we emphasize that the choice of $k$ primarily determines the black 
widow orbital evolution (see also King et al. 2005). That is, the thermal bloating of the 
companion governs the orbital evolution, while propeller and ablation evolution facilitates 
the transformation of an LMXB into a MSP binary.

\begin{figure}[t]
%\epsscale{.90}
%%\plotone{Fig6}
%\plotone{pop58test2bp}
%\includegraphics{pop58test2bp}
\includegraphics[width=84mm]{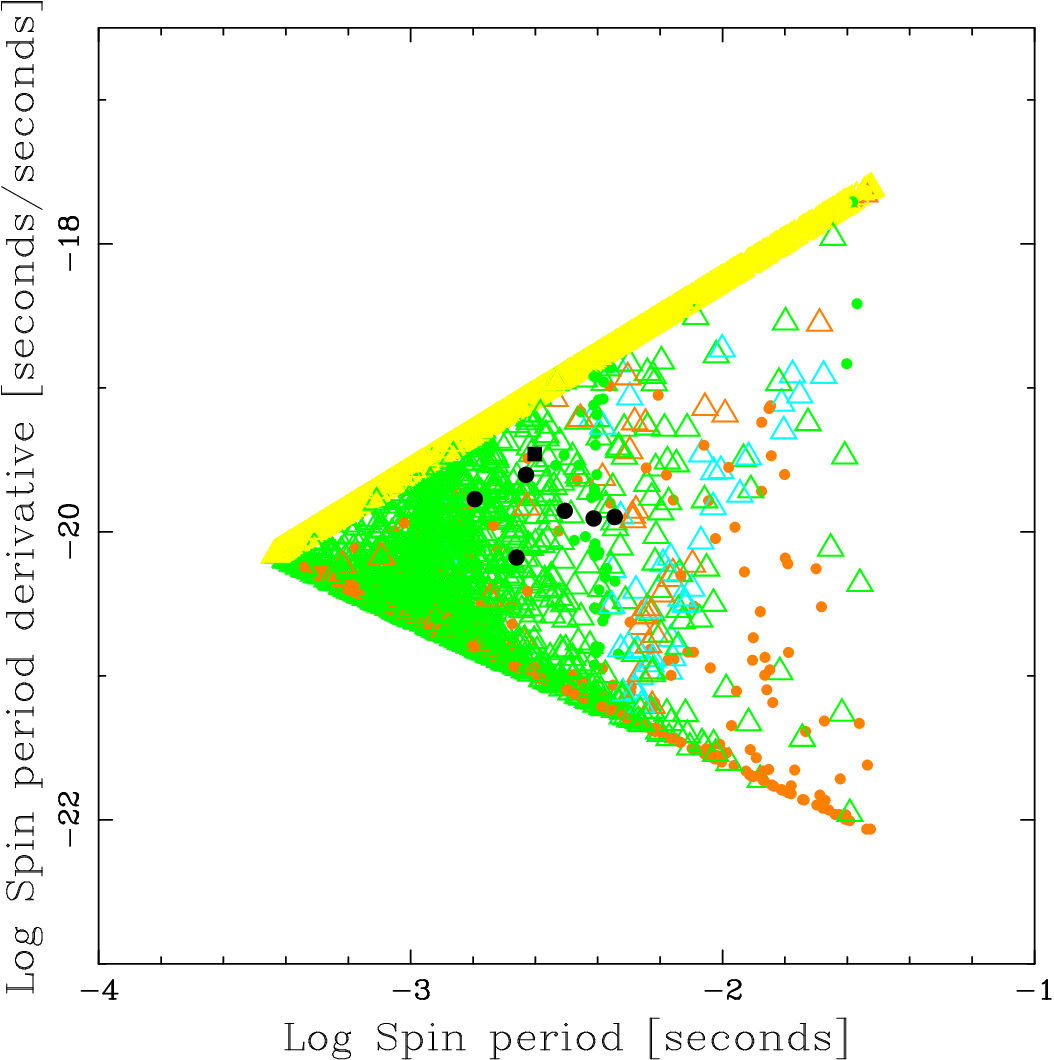}
\caption{The spin period-spin period derivative of our model B systems depicted in Fig.~\ref{f:Fig1}.  The 
model points and colors are those of Fig.~\ref{f:Fig1}, model B, where we $k=2$, $f_{\rm e}=0.001$ ($0.1\%$) 
and $f_{\rm eq}=0.5$.  Overlaid on our model are six black widow pulsars with known $\dot{P}$.  The 
black square is J1808 (Watts et al. 2008).}
\label{f:Fig6}
\end{figure}

As noted above it is evident that the longer orbital periods obtained in the present work, compared 
to the models of Lin et al. (2011), are a direct consequence of the value of $k$ adopted.  This may 
suggest that mechanisms not considered for driving the donor further out of thermal equilibrium are 
required (see below).  Provided that such a mechanism exists, we further explore the consequences of 
our model results on the properties of this population.  In particular, we find that there are three 
distinct regions of interest in $P\dot{P}$ phase space in model B as shown in Fig.~\ref{f:Fig6}.  
The lower branch represents systems characterized by an asymptotic magnetic field assumed as 
$B_{\rm bot} = 5\times10^7~{\rm G}$ (see Kiel et al. 2008).  The accumulation of systems in the 
vicinity of this branch reflects the fact that the magnetic field decays slowly with the accretion 
of mass as the field approaches this asymptotic value.  The near vertical line at ${\rm P} = 
0.0016~{\rm s}$ consists of accreting WD-MSP systems and reflects the angular momentum 
accretion timescale increasing with shorter spin periods.  
The third branch, corresponding to large spin period derivatives, represents the equilibrium 
spin up line.
Accumulation of systems on this line can occur at slower spin periods if we decrease the mass 
transfer rate, as this evolution is driven by timescales, although spin up and field decay occur 
slower as less mass is accreted.  Here, the binary evolution is dominated by accretion and the 
action of the propeller mechanism.  In this model we limit the minimum pulsar spin period to its 
mass shedding limit, a value  ${\rm P}_{\rm s} \sim 0.00025~{\rm s}$.  The MS-MSP systems that do 
not lie on the equilibrium line are characterized by a shorter angular momentum accretion timescale, 
reflecting the differing progenitor systems, as compared to their WD-MSP counterparts.  Hence, 
they `flow' from right to left through the near vertical WD-MSP line towards the asymptotic field 
limit.  

\begin{figure}[t]
%\epsscale{.90}
%%\plotone{Fig7}
%\plotone{pop58porbfreqtest2}
%\includegraphics{pop58porbfreqtest2}
\includegraphics[width=84mm]{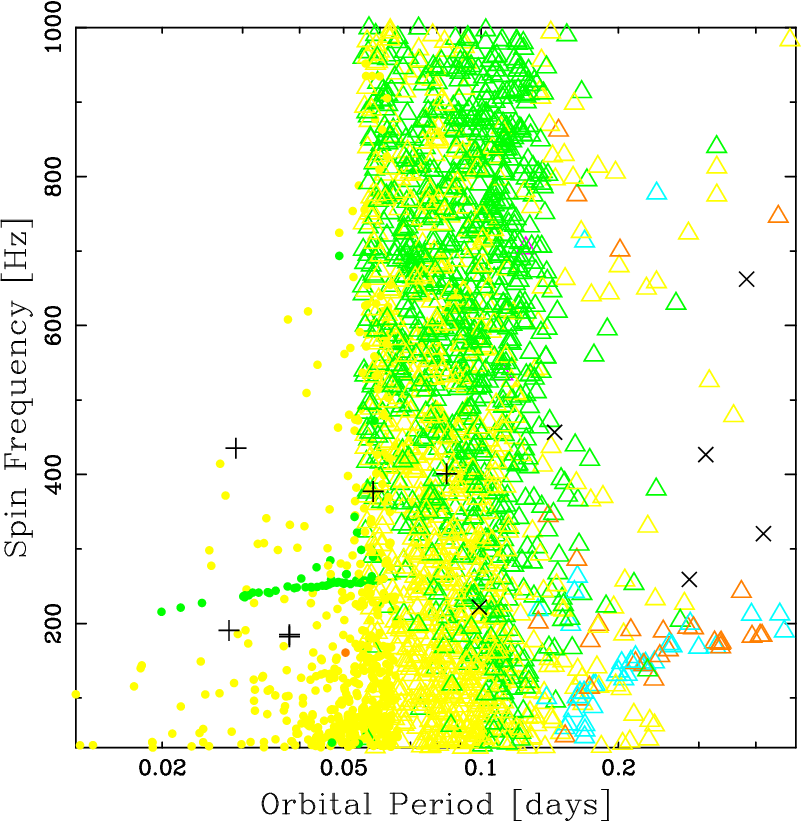}
\caption{Spin frequency vs. orbital period, comparing the results of systems from model B (colors and symbols 
as in Fig.~\ref{f:Fig1}) to observations of accretion powered MSPs (plus symbols) and rotation powered MSPs 
(cross symbols).}
\label{f:Fig7}
\end{figure}

The numerical results of model B also reveal that the ablating systems reside near the observed 
black widow systems in the spin frequency/orbital period plane (see Fig.~\ref{f:Fig7}). 
Although the entire observed range is covered in both orbital period and spin frequency, the systems in the 
high density region of our model are spinning too slowly. 
The LMXB's without ablation lie at shorter orbital periods.  The fact that accreting systems with MS 
companions on the spin equilibrium line typically have shorter orbital periods than those undergoing 
ablation highlights the importance of $\dot{M}$ and the pulsar properties in MSP formation while on 
the degenerate (or partially degenerate) tracks.  

\section{Black widow population synthesis results: formation rates and predicted numbers}
\label{s:rates}

The formation rates of the systems of interest are calculated following the methodology of 
Belczynski et al. (2002) and employed by Voss \& Tauris (2003), Pfahl et al. (2005) and Kiel, 
Hurley \& Bailes (2010). The rate is the ratio of the number of systems produced ($M_{\rm N}$) 
over the number of Type II SN produced within the model ($M_{\rm SNII}$).
This ratio is normalised to the observed Galactic Type II SN rate as estimated empirically, $R_{\rm SNII} = 
0.01~{\rm yr}^{-1}$ (Cappellaro, Evans \& Turatto 1999).  Taking into account the fraction 
of binaries, $f_{\rm B}$, the rate equation is,
\begin{equation}
R=f_{\rm B}R_{\rm SNII}\frac{M_{\rm N}}{M_{\rm NSII}}.
\end{equation}

The estimated values of our model formation rates are shown in Table~\ref{t:rates}. It can be 
seen that all the black widow birth rates reside within a narrow range between 
$4.8\times10^{-8}-3.3\times10^{-7}~{\rm yr}^{-1}$.  Decreasing the lifetime of pulsars by setting 
$\tau_{\rm B} = 100~{\rm Myr}$ (Model C) produces the greatest effect on the birth rate of 
black widow systems by causing many of the pulsars' magnetic fields to decay to the assumed 
bottom field value of $5\times10^7~$G.  That is, the particle acceleration mechanism is prevented 
and the evolution is described by continuous accretion. 
Decreasing the common envelope efficiency for a fixed stellar structure parameter (Model D) 
halves the birth rate and number of black widow systems estimated within the Galaxy, as compared 
to Model B.  Setting $k=1$ (Model E) reduces the birth rate slightly, but more 
significantly decreases the average black widow age and reducing the estimated number of 
black widows.

Assuming the companion mass function is flat in mass ratio (Model F) 
increases the average mass ratio value, causing a decrease in the number of LMXBs
and therefore black widow systems, consistent with the results of Willems \& Kolb (2002) 
and Hurley et al. (2010).  The average age of the black widow systems also decreased by a factor 
of $2$ to $\sim250~$Myr, as compared to Model B.  
Using the older mass transfer 
rate form (Model G), which is a lower rate than used in Model B, leads to an increase 
in the birth rate of black widow systems with respect to Model B and their LMXB counterparts.  
This lower rate of mass transfer falls below the critical 
values that define black widow formation (Equations~\ref{e:gammamdot}--\ref{e:lasota3}) more readily than 
those LMXBs of Model B.
Model H, where $M_\star = 1\times 10^{-6}$, highlights the importance of the magnetic field strength in
forming black widow systems.
In Section 3 we showed that assuming such a low value of $M_\star$ arrested ablation of the 
companion by the pulsar and this is the case here. 
The birth rates of black widow systems in Model H decreased substantially compared to Model B
and we find that only 5 black widow systems would be observable.

Because we examine only a subset of the binary MSP population the black widow rates are much less 
than previous binary MSP estimations which, instead, considered the total binary MSP population and 
ranged from $1\times10^{-6}-4\times10^{-4}~{\rm yr}^{-1}$ (Lorimer 1995; Cordes \& Chernoff 1997; 
Lorimer 2008; Hurley et al. 2010).

The original binary MSP estimates led to the binary MSP birth rate problem in which the model estimates 
of the MSP birth rates were higher by up to 2 orders of magnitude than estimated for LMXB birth 
rates (as initially discussed using semi-empirical arguments by Kulkarni \& Narayan 1988).
These estimates placed the LMXB binary birth rate at $\sim 10^{-7}~{\rm yr}^{-1}$ assuming 
an observable X-ray lifetime of $\sim10^9~{\rm yr}$ (Pfahl, Rappaport \& Podsiadlowski 2003; Hurley 
et al. 2010).

We show that the LMXB rates given in Table~\ref{t:rates} are more consistent with the
black widow birth rates.  However, the inconsistency may still exist from the wider binary MSP 
population for 
which the formation scenarios that exist 
vary greatly and do not necessarily require a long lived LMXB phase to form.
The fundamental reason for our better consistency between birth rates is simply that
we are examining two populations where one forms directly (and solely) from the other.

However, the necessity for thermal bloating in the models required for matching the observed 
orbital period-companion mass distribution may provide another possible reason of the overestimate 
of the LMXB age of $10^9$yr (along with irradiation induced 
mass transfer cycles suggested by Pfahl, Rappaport \& Podsiadlowski 2003).
We suggest that the thermal bloating on the degenerate track is a result 
of evolution on time scales shorter than the thermal timescale of the donor and, hence, 
the ages of LMXBs will be overestimated if the evolution is assumed to proceed on the thermal 
timescale.
If the time spent in the X-ray binary phase is less than $10^9~$yr, and account is taken of 
the duty cycle of accretion activity, the birth rate problem can be somewhat alleviated.

\begin{table*}
\small
% \centering
% \scriptsize
% \begin{minipage}{140mm}
  \caption{Characteristics of the main set of models used in this work and 
the resultant formation rates
and estimated number of black widows in the Galaxy now.  
Here $R_{\rm SNII} = 0.01~{\rm yr}^{-1}$ (Cappellaro, Evans \& Turatto 1999) and $f_{\rm B} = 0.5$.
We vary models to have different pulsar field decay timescales, $\tau_{\rm B}$, companion thermal bloating, $k$,
field decay parameter, $M_\star$, mass transfer rate, mass ratio distribution and 
common envelope $\alpha_{\rm CE}$ and $\lambda$.
We include model A for completeness, however, without propeller effects or ablation of 
the companion, it does not produce black widow systems.
  \label{t:rates}}
%  \begin{tabular}{crrrrrrrr}
  \begin{tabular}{@{}crrrrrrrr@{}}
%  \hline
\tableline
 Model & $\tau_{\rm B}$ [Myr] & $k$ & $M_\star$ & MT method & MRD & $\alpha_{\rm CE},~\lambda$ & Birth rates $[{\rm yr}^{-1}]$ & Number \\
   &  &  &  &  &  &  & LMXB::MSP  &  \\
\hline 
 A & $900$ & $2$ & $1e-4$ & Eq.~\ref{e:newmdot} & KTG93 & $3$, variable & $2.2e-5::N/A$ & N/A \\
 B & $900$ & $2$ & $1e-4$ & Eq.~\ref{e:newmdot} & KTG93 & $3$, variable & $1.4e-5::2.2e-7$ & $110$ \\
 C & $100$ & $2$ & $1e-4$ & Eq.~\ref{e:newmdot} & KTG93 & $3$, variable & $1.8e-5::5.0e-8$ & $13$ \\
 D & $900$ & $2$ & $1e-4$ & Eq.~\ref{e:newmdot} & KTG93 & $1$, $0.5$ & $6.3e-6::1.0e-7$ & $50$ \\
 E & $900$ & $1$ & $1e-4$ & Eq.~\ref{e:newmdot} & KTG93 & $3$, variable & $1.4e-5::1.0e-7$ & $25$ \\
 F & $900$ & $2$ & $1e-4$ & Eq.~\ref{e:newmdot} & flat in $q$ & $3$, variable & $6.2e-6::6.0e-8$ & $13$ \\
 G & $900$ & $2$ & $1e-4$ & Eq.~\ref{e:bsemdot} & KTG93 & $3$, variable & $1.2e-5::3.3e-7$ & $179$ \\
 H & $900$ & $2$ & $1e-6$ & Eq.~\ref{e:newmdot} & KTG93 & $3$, variable & $2.0e-5::4.8e-8$ & $5$ \\
%\hline
\tableline
\end{tabular}
\end{table*}

\section{Discussion} 

The results of our simulations (Model B) reveal that systems can be produced with orbital periods as 
long as 0.4 days and with companion masses as low as $0.005 \msun$, with the orbital period depending 
upon the thermal bloating factor. In contrast, without thermal bloating ($k=1$) systems form with 
masses as low as $\sim 0.002\msun$ and orbital periods $\gtrsim 0.15~$days before they are destroyed. 
In general, our model systems become rotation powered MSPs in a similar region of the orbital 
period/companion mass plane where the observed LMXBs apparently turn into radio MSPs. 
We find a dependence on the bloating factor for the orbital period at which the transition from 
the LMXB phase to MSP occurs, however, this dependence is mitigated somewhat if the system 
undergoes the transition prior to the companion reaching a sufficiently low mass to become 
semi-degenerate (especially for $k=1$).  This transition typically occurs at $P_{\rm orb} \sim 0.1~$ 
days.  Without the effects of propeller action and ablation (see model A), the mass accretion 
varies continuously and the evolution to the rotation powered MSP phase is inhibited.  This latter 
model is similar to the models of Hurley et al. (2010), which well describes the observed LMXB 
population along the UCXB WD branch (e.g., SWIFT J1756.9-2508) with lower mass companions and the 
MS branch (e.g., SAX J1808.4-3658) for higher mass companions.  However, the MS branch in models 
without propeller action and ablation does not extend to sufficiently low masses to describe many 
observed systems (e.g., J2241-5236, which has $P_{\rm orb} = 0.146~$days and a minimum companion 
mass of $0.012\msun$; Keith et al. 2011), assuming that thermal bloating is important. 

The systems that are regularly accreting while on the equilibrium line will be more difficult 
to detect as rotation powered MSPs, either because accretion precludes the activation of particle 
acceleration mechanism in the pulsar magnetosphere or because the pulsed emission is heavily 
obscured by the ejected material.  This evolutionary phase is particularly dominant at very low 
companion mass, $M_{\rm d} < 0.016\msun$, and is, perhaps, the reason for the observed lack of 
systems in this region of parameter space (although lower masses lead to lower ablated material 
surrounding the system).  In addition, the possible occurrence of intermittent accretion during the 
propeller and spin equilibrium phase, as shown in numerical models of Romanova et al. (2009), 
may further hinder radio MSP detection.

The observations and numerical results reveal that the ablating companions are only present on the 
low mass semi-degenerate MS branch and absent on the UCXB branch.  The lack of black widow pulsars 
on the latter branch (with shorter orbital periods) may reflect the higher mass transfer rates driven 
by the greater losses of orbital angular momentum associated with gravitational radiation 
characteristic of systems on this branch.  This property can limit the degree of instability in the 
accretion disk, thereby, restricting the parameter range for the effectiveness of ablation.  
It is interesting to note that the WD-MSP branch approaches an orbital period of $\sim 0.1~{\rm day}$ 
which is where we find a transition from the X-ray binary to the MSP binary phase. This 
suggests that systems like HETE J1900.1-2455 with $P_{\rm orb}\sim 0.058~{\rm days}$ and $M_{\rm c} 
\sim 0.018 \msun$ (Kaaret et al. 2006) may eventually become MSP binaries.  We note that HETE J1900.1 
may in fact contain a hybrid (He-rich) WD companion which would result in a companion lying between 
the MS-WD branches, which can be seen in Fig~\ref{f:Fig1} where HETE J1900.1 lies just below the 
$k=1$ model.  Such an evolution could have involved companions with initial masses greater than about 
$1 \msun$ in which nuclear evolution has occurred (e.g., Nelson \& Rappaport 2003).  At present 
\textsc{bse} does not model or evolve such hybrid stars.

It has been argued by Ho, Maccarone \& Andersson (2011) that the location of MSP's in the 
spin frequency/orbital period plane could be understood in terms of the comparisons of orbital 
period changes, spin up and spin down timescales.  Specifically, their location depends on where 
spin up/down and orbital period increase/decrease dominate the evolution of the orbit and MSP spin.  
In contrast to the description adopted here they emphasize the importance of gravitational wave spin 
down of accreting NSs.  However, we find an evolution including propeller and ablation physics during 
the LMXB phase naturally leads to the absence of accreting MSPs at long orbital periods ($> 0.1$ day) 
and low spin frequencies ($< 600$ Hz) without the necessity for the inclusion of gravitational wave 
emission from a rapidly rotating NS.

\section{Conclusion}

A model of the formation and evolution of radio MSPs has been carried out using a binary 
population synthesis method (the \textsc{bse} code) incorporating the effects of mass ejection 
from the system associated with propeller action and ablation. The accuracy of \textsc{bse} to 
evolve LMXBs with NSs as the compact object was assessed by comparison to detailed codes of McDermott 
\& Taam (1989) and Lin et al. (2011). As a first step in quantifying the consequences of 
including new input physics related to the propeller phase and ablation of the NS companion on 
the binary evolution, the population synthesis method was adopted to explore a large parameter range 
and number of systems.

It has been shown that by including both propeller and ablation physics into our models of binary 
and stellar evolution the companion to the neutron star in the binary system can detach from 
its Roche lobe transforming a LMXB to a rotation powered MSP binary.  In particular, it is found 
that the operation of the propeller action is as crucial in this transition as the ablation mechanism.
Our work also highlights the importance of thermal bloating of the companion star beyond that 
found in previous detailed stellar models as it relates to the orbital periods of black widow pulsar 
systems (see also King et al 2005).

Although the evolutionary model is promising, our model MSPs are characterized by magnetic fields 
that are high ($\sim 8\times 10^8~{\rm G}$) as compared to the observed systems ($\sim 2\times 
10^8~{\rm G}$).  Since this is determined by our simple spin equilibrium line, modifications to 
bring the model results into closer agreement with observations will be necessary.  This may also 
be related to the fact that the majority of our black widow systems are spinning slower than 
observations suggest.

Finally, the comparison of our model results with observations suggest that the degree of thermal 
bloating as parameterized by the factor $k$ is $\sim 2$.  The magnitude of this effect may suggest 
that additional processes driving the mass transfer in these systems may be operating during the 
LMXB phase.  Examples of such processes include the influence of X-ray irradiation induced stellar 
winds as proposed by Podsiadlowski (1991) and Iben et al. (1997) or tidal heating as hypothesized by 
Rasio et al. (2000).  Including such effects, where it is likely that the degree of thermal bloating 
varies, and following the binary evolution during the LMXB phase to the binary radio millisecond 
pulsar phase for incorporation into detailed stellar structure and evolution models is highly 
desirable.

\acknowledgements 
We thank the referee for his/her comments which have improved the clarity and presentation of this 
paper.  This work was supported in part by the Theoretical Institute for Advanced Research in 
Astrophysics (TIARA) operated under the Academia Sinica Institute of Astronomy \& Astrophysics in 
Taipei, Taiwan and by NASA ATP Grants NNX09AO36G and NNX08AG66G and an NSF AST-0703950 to 
Northwestern University.  We also thank Swinburne University of Technology for use of `the 
Green machine', the supercomputer on which the simulations were completed.  PDK thanks the 
Academia Sinica Institute of Astronomy \& Astrophysics for their hospitality during his visit.

%\begin{thebibliography}{}
\bf{References}

Andronov, N., Pinsonneault, M., \& Sills, A. 2003, \apj, 582, 358 \\
Archibald, A. M., et al. 2009, Science, 324, 1411 \\
Arzoumanian, Z., Cordes, J. M., \& Wasserman, I. 1999, \apj, 520, 696 \\
Barr, E. D., et al., 2013, \mnras, 429, 1633 \\
Bates, S. D., et al. 2011, MNRAS, 416, 2455 \\
Belczynski K., Kalogera V., Bulik T., 2002, \apj, 572, 407 \\
Belczynski K., Kalogera V., Rasio F. A., Taam R. E., Zezas A., Bulik T.,
Maccarone T. J., Ivanova N., (2008), ApJS, 174, 223 \\
%\bibitem[]{}2012MNRAS.427L..90R
Bodenheimer, P., \& Taam, R. E. 1984, ApJ, 280, 771 \\
Bond, H. E., White, R. L., Becker, R. H., \& OÕBrien, M. S. 2002, PASP, 114, 1359 \\
Boyles, J., et al. 2011, AIPS, 1357, 32 \\
Burgay, M., et al., 2006, MNRAS, 368, 283 \\
Cappellaro, E., Evans, R., Turatto, M., 1999, A\&A, 351, 459 \\
Chaboyer, B., Demarque, P., \& Pinsonneault, M. H. 1995, \apj, 441, 865 \\
Cordes, J. M., Chernoff, D. F., 1997, \apj. 482, 971 \\
de Kool M., 1990, \apj, 358, 189 \\
Dewi J. D. M., Tauris T. M., 2001, in Podsiadlowski P., Rappaport S., King
A. R., DÕAntona F., Burderi L., eds, ASP Conf. Ser. Vol. 229, Evolution of Binary and Multiple Star Systems. Astron. Soc. Pac., San Francisco, p. 255 \\
%\bibitem[]{E83} Eggleton, P. P. 1983, \apj, 268, 368
Frank, J., King, A., \& Raine, D.J. 2002, Accretion Power in Astrophysics (Cambridge: Cambridge Univ. Press)
Fruchter, A. S., Stinebring, D. R. \& Taylor, J. H., 1988, Nature, 333, 237 \\
Gentile, P., et al., 2012, arXiv:1210.7342G \\
Hessels, J. W. T., et al. 2011, to appear in AIP Conference Proceedings of Pulsar Conference 2010 "Radio Pulsars: a key to unlock the secrets of the Universe", Sardinia, October 2010, ed. Burgay, M., D'Amico, N., Esposito, P., Pellizzoni, A., Possenti, A. \\
Ho, W. C. G., Maccarone, T. J., \& Nils, A. 2011, \apj, 730, 36 \\
Hurley, J. R., Pols, O. R., \& Tout, C. A. 2000, \mnras, 315, 543 \\
Hurley, J. R., Tout, C. A., \& Pols, O. R. 2002, \mnras, 329, 897 \\
Hurley, J. R., Tout, C. A., Wickramasinghe, D. T., Ferrario, L., Kiel, P. D. 2010, \mnras, 406, 656 \\
Iben I. Jr., Livio M., 1993, PASP, 105, 1373 \\
Iben, Jr. I., Tutukov, A. V., \& Fedorova, A. V. 1997, \apj, 486, 955 \\
Illarionov, A. F., \& Sunyaev, R. A. 1975, A\&A, 39, 185 \\
Ivanova, N, \& Taam, R. E. 2003, \apj, 599, 516 \\
Kaaret, P., Morgen, E. H., Vanderspek, R., \& Tomsick, J. A. 2006, \apj, 638, 963 \\
Keith, M. J., et al. 2011, \mnras, 414, 1292 \\
Keith, M. J., Johnston, S., Bailes, M., et al. 2012, MNRAS, 419, 1752 \\
Kerr, M., et al. 2012, \apj, 748 , 2 \\
Kiel, P. D., \& Hurley, J. R. 2006, \mnras, 369, 1152 \\
Kiel, P. D., Hurley, J. R., Bailes, M., \& Murray, J. R. 2008, \mnras, 388, 393 \\
Kiel, P. D., Hurley, J. R., \& Bailes, M. 2010, \mnras, 406, 656 \\
King, A. R., Beer, M. E., Rolfe, D. J., Schenker, K., \& Skipp, J. M. 2005, \mnras, 385, 1501 \\
Kulkarni S.R., Narayan R., 1988, \apj, 335, 755 \\
Lasota, J.-P., Dubus, G., \& Kruk, K. 2008, A\&A, 486, 523 \\
Lin, J., Rappaport, S., Podsiadlowski, P., Nelson, L.;,Paxton, B., Todorov, P., 2011, \apj, 732, 70 \\
Lorimer, D. C., 1995, \mnras, 274, 300 \\
Lorimer, D. L., 2008, LRR, 11, 8 \\
Manchester, R. N., Hobbs, G. B., Teoh, A., Hobbs, M. 2005, \apj, 129, 1993 \\
McDermott, P. N., \& Taam, R. E. 1989, \apj, 342, 1019 \\
Nelemans G., \& Tout C. A., 2005, \mnras, 356, 753 \\
Nelson, L. A., \& Rappaport, S. 2003, \apj, 598, 431 \\
Ostriker, J. P. \& Gunn, J. E., 1969, \apj, 157, 1395 \\
Paczynski, B., 1976, in Eggleton, P., Milton, S., Whelan, J., eds, 
Proc. IAU Symp. 73, Structure and Evolution of Close Binary Systems. Reidel, Dordrecht, p. 75
Paczynski, B., \& Sienkiewicz, R., 1981, \apj, 248, L27 \\
Pfahl E., Rappaport S., Podsiadlowski P., 2003, \apj, 597, 1036 \\
Pfahl E., Podsiadlowski P., Rappaport S., 2005, \apj, 628, 343 \\
Phinney, E. S., Evans, C. R., Blandford, R. D., \& Kulkarni, S. R. 1988, Nature, 333, 832 \\
Podsiadlowski, P. 1991, \nat, 350, 136 \\
Podsiadlowski P., Rappaport S., Han Z., 2003, \mnras, 341, 385 \\
Pols O. R., Schroéder K. P., Hurley J. R., Tout C. A., Eggleton P. P., 1998,
\mnras, 298, 525 \\
Podsiadlowski, P., Ivanova, N., Justham, S., Rappaport, S., 2010, \mnras, 406, 840 \\
Portegies Zwart S. F., Yungelson A., 1998, A\&A, 372, 173 \\
Ransom, S. M., et al. 2011, \apj, 727, 16 \\
Rappaport, S., Joss, P. C., \& Webbink, R. F. 1982, \apj, 254, 616 \\
Rasio, F., Pfahl, E. D., \& Rappaport, S. 2000, \apj, 532, L47 \\
Ray, P. S., et al. 2012, arXiv:1205.3089 \\
Ricker, P. M., Taam, R., 2008, \apj, 672, 41 \\
Ritter, H., \& Kolb, U. 2003, A\&A, 389, 485 \\
Roberts, M. S. E. 2011,  AIP Conference Proceedings of Pulsar Conference 2010 "Radio Pulsars: a key to unlock the secrets of the Universe", Sardinia, October 2010, ed. Burgay, M., D'Amico, N., Esposito, P., Pellizzoni, A., Possenti, A.
Roberts, M. S. E. 2012,  to appear in IAU Conference Proceedings "Neutron Stars and Pulsars: Challenges and Opportunities after 80 years", ed. van Leeuwen, J., arXiv:1210.6903 \\
Romanova, M. M., Ustyugova, G. V., Koldoba, A. V., \& Lovelace, R. V. E. 2009,  
\mnras, 399, 1802 \\
Ruderman, M. A., \& Shaham, J. 1985, \apj, 289, 244 \\
Ruderman, M. A., Shaham, J., \& Tavani, M. 1989, \apj, 336, 507 \\
Schatzman, E., 1962, Ann. Astrophys, 25, 18 \\
Shibazaki, N., Murakami, T., Shaham, T., Nomoto, K., 1989, Nature, 342, 656 \\
Stappers, et al., 1996, \apj, 465, L119 \\
Takata, J., Cheng, K. S., \& Taam, R. E. 2010, \apj, 723, 68 \\
Taam, R. E., \& Sandquist, E. L. 2000, ARA\&A, 38, 111 \\
Thorstensen, J. R., \& Armstrong, E. 2005, AJ, 130, 759 \\
Tout, C. A., Aarseth, S. J., Pols, O. R., \& Eggleton, P. P. 1997, \mnras, 291, 732 \\
Tutukov A., Yungelson L., 1996, \mnras, 280, 1035 \\
van den Heuvel E. P. J., \& van Paradijs, J. 1988, Nature, 334, 227 \\
Verbunt, F., \& Zwaan, C. 1981, A\&A, 100, L7 \\
Voss R., Tauris T. M., 2003, \mnras, 342, 1169 \\
Wang, Z., Archibald, A. M., Thorstensen, J. R., Kaspi, V. M., Lorimer, D. R., 
Stairs, I., \& Ransom, S. M. 2009, \apj, 703, 2017 \\
Watts, A. L., Krishnan, B., Bildsten, L., \& Shutz, B. F. 2008, \mnras, 389, 839 \\
Webbink R. F., 1984, \apj, 277, 355 \\
Willems B., Kolb U., 2002, \mnras, 337, 1004 \\
Woudt, P. A., Warner, B., \& Pretorius, M. L. 2004, \mnras, 351, 1015 \\
Yungelson L. R., Livio M., Tutukov A. V., Saffer R. A., 1994, \apj, 420, 336 \\
%\end{thebibliography}
\end{document}